\documentclass[useAMS,usenatbib]{mn2e}
\usepackage{graphicx}
\usepackage{amsmath,amssymb}
\usepackage{hyperref}
\usepackage{enumitem}
\usepackage{color}
\usepackage{xcolor}
\usepackage{tabularx}

\newcommand\apj{ApJ}
\newcommand\apjl{ApJL}
\newcommand\apjs{ApJS}
\newcommand\mnras{MNRAS}
\newcommand\aap{A\&A}
\newcommand\aaps{A\&AS}

\newcommand\pasa{PASA}
\newcommand\pasj{PASJ}

\newcommand\prd{Phys. Rev. D}
\newcommand\prc{Phys. Rev. C}

\newcommand\physrep{Physics Reports}
\newcommand{\jqsrt}{J.~Quant.~Spec.~Radiat.~Transf.}

\newcommand{\dd}{{\rm d}}

\newcommand{\eps}{\epsilon}
\newcommand{\vecF}{\textbf{\em F}}
\newcommand{\vecn}{\textbf{\em n}}
\newcommand{\Msol}{M_{\odot}}
\newcommand{\tpb}{t_{\textrm{pb}}}
\newcommand{\n}{\mathsf{n}}

\def\ga{\,\,\raise0.14em\hbox{$>$}\kern-0.76em\lower0.28em\hbox
{$\sim$}\,\,}
\def\la{\,\,\raise0.14em\hbox{$<$}\kern-0.76em\lower0.28em\hbox
{$\sim$}\,\,}

\title[Comparison of CCSN codes \& approximations]
{Core-collapse supernova simulations in one and two dimensions: comparison of codes and approximations}
\author[Just et al.]{O.~Just$^{1,2}$\thanks{oliver.just@riken.jp}, R.~Bollig$^{2,3}$, H.-Th.~Janka$^{2}$, M.~Obergaulinger$^4$, R.~Glas$^{2,3}$, \newauthor  \& S.~Nagataki$^{1,5}$   \\
  $^1$ Astrophysical Big Bang Laboratory, RIKEN Cluster for Pioneering Research, 2-1 Hirosawa, Wako, Saitama 351-0198, Japan \\
  $^2$ Max-Planck-Institut f\"ur Astrophysik, Karl-Schwarzschild-Str. 1, 85748 Garching, Germany \\
  $^3$ Physik Department, Technische Universit\"at M\"unchen, James-Franck-Stra{\ss}e 1, 85748 Garching, Germany \\
  $^4$ Departament d{\'{}}Astronomia i Astrof{\'i}sica, Universitat de Val{\`e}ncia,
  Edifici d{\'{}}Investigaci{\'o} Jeroni Mu{\~n}oz, \\
  C/ Dr.~Moliner, 50, E-46100 Burjassot (Val{\`e}ncia), Spain \\
  $^5$ RIKEN Interdisciplinary Theoretical \& Mathematical Science Program (iTHEMS), 2-1 Hirosawa, Wako, Saitama, Japan 351-0198  
}

\begin{document}

\maketitle

\label{firstpage}

\begin{abstract}
We present spherically symmetric (1D) and axisymmetric (2D) supernova simulations for a convection-dominated 9\,$M_\odot$ and a 20\,$M_\odot$ progenitor that develops violent activity by the standing-accretion-shock instability (SASI). We compare in detail the \textsc{Aenus-Alcar} code, which uses fully multidimensional two-moment neutrino transport with an M1 closure, with a ray-by-ray-plus (RbR+) version of this code and with the \textsc{Prometheus-Vertex} code that employs RbR+ two-moment transport with a Boltzmann closure. Besides testing consequences of ignored non-radial neutrino-flux components in the RbR+ approximation, we also discuss the influence of various transport ingredients applied or not applied in recent literature, namely simplified neutrino-pair processes, neutrino-electron scattering, velocity-dependent and gravitational-redshift terms, and strangeness and many-body corrections for neutrino-nucleon scattering. \textsc{Alcar} and \textsc{Vertex} show excellent agreement in 1D and 2D despite a slightly but systematically smaller radius ($\sim$1\,km) and stronger convection of the proto-neutron star with \textsc{Alcar}. As found previously, the RbR+ approximation is conducive to explosions, but much less severely in the convection-dominated 9\,$M_\odot$ case than in the marginally exploding 20\,$M_\odot$ model, where the onset time of explosion also exhibits big stochastic variations, and the RbR+ approximation has no distinctly stronger supportive effect than simplified pair processes or strangeness and many-body corrections. Neglecting neutrino-electron scattering has clearly unfavorable effects for explosions, while ignoring velocity and gravitational-redshift effects can both promote or delay the explosion. The ratio of advection timescale to neutrino-heating timescale in 1D simulations is a sensitive indicator of the influence of physics ingredients on explosions also in multidimensional simulations.
\end{abstract}


\section{Introduction}\label{sec:introduction}

Rarely any astrophysical phenomenon has been studied as long and as intensely using the most powerful available supercomputers as the explosion mechanism of core-collapse supernovae \citep[CCSNe; see, e.g.,][for recent reviews]{Foglizzo2015a,Janka2016a,Muller2016a}. After several decades of research the neutrino-driven mechanism, first suggested by \citet{Colgate1966} and later in its modern version by \citet{Bethe1985}, remains the leading candidate responsible for the revival of the initially stalled shock wave in ordinary CCSNe. Although current three-dimensional simulations lend support to the scenario that neutrino heating powers the explosion \citep{Takiwaki2014a, Melson2015a,Melson2015b, Lentz2015a, Roberts2016a, Ott2017a} important questions remain to be studied in more detail and are vividly investigated by a large number of researchers worldwide. These questions concern, e.g., the relative importance of the standing accretion-shock instability \citep[SASI;][]{Blondin2003} and post-shock convection \citep[e.g.][]{Fernandez2015a}, the impact of improved neutrino interactions \citep[e.g.][]{Kotake2018a}, the role of turbulence \citep[e.g.][]{Radice2018a}, the dependence on the progenitor \citep[e.g.][]{Sukhbold2016a} and its pre-collapse asymmetries \citep[e.g.][]{Muller2017a}, the quest for a universal explosion criterion \citep[e.g.][]{Ertl2016a}, the information contained in the emitted neutrinos \citep[e.g.][]{Tamborra2017a}, gravitational waves \citep[e.g.][]{Richers2017b}, and remnant structure \citep[e.g.][]{Ono2013a}, or the role of rotation and magnetic fields
\citep[e.g.][]{Obergaulinger2017}.

All these research lines depend in some way or another on results obtained from time-dependent models of the supernova core, which are provided by multidimensional (neutrino-) hydrodynamics simulations. One of the major obstacles for detailed CCSN models is the neutrino transport. Ideally, the latter requires to solve along with the three-dimensional hydrodynamics equations a Boltzmann equation depending on six phase-space coordinates. However, since this is not possible with present supercomputer capabilities, at least not for reasonable resolution and number of time steps, various schemes and approximations have been proposed in the past, which differ considerably concerning their computational efficiency and accuracy.

The computationally most efficient methods of handling neutrino processes are light-bulb \citep[e.g.][]{Hanke2012a,Couch2015} and leakage \citep{Ruffert1996a,OConnor2010a,Perego2016a} schemes, which add almost no cost to the hydrodynamics equations since the neutrino source terms entering the latter are estimated based on the hydro-/thermodynamic information only. More involved schemes, such as the fast multigroup transport \citep{Muller2015a} or the one designed by \citet{Scheck2008}, solve the conservation equation for neutrino energy assuming a pre-defined profile for the ratio of the flux density to energy density. Another scheme with relatively high computational efficiency is the isotropic diffusion source approximation \citep[IDSA;][]{Liebendorfer2009,Takiwaki2014a,Suwa2016b} that decomposes the neutrino distribution into a trapped and a free-streaming component, which are separately evolved using a parabolic diffusion equation and an elliptic equation, respectively. One of the most often used schemes, also outside of CCSN theory, is flux-limited diffusion \citep[FLD;][]{Levermore1981,Dessart2006,Bruenn2016a,Dolence2015a}, which evolves the energy density, $E$, of neutrinos assuming the flux density, $F$, to be given locally by a generalized (``flux-limited'') diffusion law that limits the flux to remain causal, $F\leq c E$ ($c$ being the speed of light).

More recently, the so-called M1 scheme \citep{Minerbo1978,Levermore1984,Pons2000,Audit2002,Shibata2011} was implemented in a number of neutrino-transport codes \citep[][]{OConnor2015a,Foucart2015,Sekiguchi2015a,Just2015b,Kuroda2016,Skinner2016}. Being closely related to FLD, M1 evolves additionally to the latter the flux densities. The resulting two-moment system -- energy and flux density are the 0th- and 1st-order angular moments of the specific neutrino intensity -- is augmented with a local closure relation expressing the 2nd-order moments (i.e. the neutrino pressure tensor) as a function of the evolved moments. The M1 scheme comes with the convenient property of being hyperbolic, which, together with the fact that fluid velocities are high in CCSNe, allows to advance the equations using explicit time stepping. This, in turn, makes it computationally feasible to solve the unconstrained, fully multidimensional transport equations.

Finally, in contrast to the aforementioned approximate schemes, more involved but also considerably more expensive schemes, so-called Boltzmann solvers, attempt to resolve the full angular dependence
of the radiation field. This can be achieved, e.g. by employing tangent ray \citep{Burrows2000,Rampp2002,Buras2006}, discrete ordinate \citep{Mezzacappa1993a,Yamada1999a,Liebendorfer2004,Nagakura2018a}, or Monte Carlo methods \citep{Janka1989a,Richers2017a,Foucart2018a} to evolve the Boltzmann equation either directly or coupled to a two-moment system.

One measure to make expensive solvers more affordable in axisymmetric \citep[2D; e.g.][]{Buras2006,Muller2012b,Bruenn2016a,Nakamura2015a} and three-dimensional \citep[3D; e.g.][]{Takiwaki2014a,Melson2015a,Lentz2015a,Muller2017a,Wongwathanarat2017a} simulations is to employ the ray-by-ray(-plus) approximation (RbR+), which neglects the evolution of the non-radial flux components and therefore allows to evolve basically one-dimensional (1D) transport problems on each angular bin along the radial directions in a quasi-decoupled manner. This leads to a major boost of computational performance, particularly for a scheme where the time integration requires the inversion of matrices spanned over the entire grid, because then the total number of operations is reduced by much more than a factor of three and the code can be efficiently parallelized. The ray-by-ray approach is naturally motivated by the rather spherical geometry of the SN core and resulting subdominance of non-radial neutrino fluxes, and by the notion that local anisotropies enhanced by RbR+ could average out and thus only leave a small imprint on the angle- and time-averaged dynamics \citep{Buras2006}. On the other hand, the suspicion was raised \citep{Dolence2015a,Sumiyoshi2015a} that the RbR+ approximation could enhance the feedback between the fluid flow and neutrino radiation, which could induce explosions more readily. However, so far the RbR+ approximation was investigated self-consistently, i.e. by comparing time-dependent simulations with and without using RbR+, only using one code \citep{Skinner2016} and more detailed exploration is warranted.

In addition to the large number of available numerical approximations of the neutrino transport, probably an even larger number of (micro-)physics ingredients exists in various degrees of sophistication. For instance, while some studies use a large set of state-of-the-art neutrino interactions \citep[e.g.][]{Melson2015a,Lentz2015a,Kotake2018a}, many other studies rely on a smaller and more basic set of interactions \citep[often adopted from][]{Bruenn1985} and neglect or simplify numerically cumbersome reactions that couple different neutrino energies, such as inelastic scattering of neutrinos off electrons and positrons as well as pair processes (e.g. electron-positron annihilation and nucleon-nucleon bremsstrahlung). Similarly, frame-dependent effects such as Doppler- and gravitational energy-shifts are sometimes neglected for numerical convenience by dropping all velocity-dependent terms and energy derivatives in the evolution equations. Other features, such as strangeness \citep{Horowitz2002} or many-body \citep{Horowitz2017a} corrections to the neutral-current cross sections for neutrino-nucleon scattering can be implemented rather easily and have recently found their way into a number of simulations \citep[e.g.][]{Melson2015a,Bollig2017a,Burrows2018a,Kotake2018a}.

With the growing number of numerical models, also the demand grows to perform systematic comparisons between codes, methods and approximations. Method comparisons, such as the ones conducted by \citet{Liebendorfer2005, Lentz2012a, Lentz2012b, Muller2012b, Richers2017a, OConnor2018c, Pan2018a, Cabezon2018a}, might not only help improving the employed algorithms on all sides, e.g. by unmasking previously unknown deficiencies. They may also prove useful to other groups in locating the reasons why their codes give different results. Moreover, they may provide instrumental help in developing and gauging code extensions or new codes. Unfortunately, going beyond spherically symmetric models when comparing time-dependent CCSN simulations is significantly more challenging because of the appearance of (numerical and physical) perturbations, fluid instabilities, turbulence, and stochasticity. The last mentioned issue, the importance of which was emphasized by \citet{Cardall2015a}, may be a particular cause of complication if simulations are heavily time consuming and expensive and therefore only allow for a small number of runs to be conducted.

In this paper we use the M1-based code \textsc{Aenus-Alcar} \citep{Just2015b, Obergaulinger2008a} to compare in spherical- and axisymmetry against the well-known code \textsc{Prometheus-Vertex} \citep{Rampp2002,Buras2006} that employs the RbR+ approximation with a Boltzmann closure, and to test the ramifications of the RbR+ approximation in axisymmetry together with those of other frequently used modeling simplifications in the transport sector. More specifically, we address the following questions:

\begin{enumerate}[label=\arabic*.,leftmargin=3mm]
\item How good is the agreement between the M1 code \textsc{Alcar} and the Boltzmann solver \textsc{Vertex} in 1D given exactly the same input physics?
\item How good is the corresponding agreement in 2D?
\item What is the impact of RbR+ on the explodability and on proto-neutron star (PNS) convection, how does this impact depend on the stellar progenitor, and how does it compare to the other considered modeling variations?
\item What difference does it make if neutrino-electron scattering is neglected, pair processes are simplified, and strangeness as well as many-body corrections are included?
\item What happens if velocity-dependent terms as well as gravitational redshift are neglected in the transport?
\item What is the impact of stochasticity, e.g. how large is the scatter of explosion times if simulations are repeated several times?
\end{enumerate}

The paper is organized as follows: In Sect.~\ref{sec:comp-meth} we outline the numerical methods and describe the setup of our models, in Sects.~\ref{sec:comp-with-vert} and~\ref{sec:2d-axisymm-models} we present the results for our 1D and 2D models, respectively, and in Sect.~\ref{sec:summary-conclusions} we summarize and conclude.

\section{Numerical methods and model setup}\label{sec:comp-meth}

\subsection{Governing equations and discretization schemes}

In the following we outline the main features of the employed simulation tools, \textsc{Alcar} and \textsc{Vertex}. For in-depth descriptions of these codes we refer the reader to \citet{Just2015b} as well as \citet{Rampp2002} and \citet{Buras2006}, respectively.

Both codes employ a Godunov-type finite-volume scheme in spherical polar coordinates to solve the equations of Newtonian hydrodynamics with an effective general relativistic gravitational potential \citep{Marek2006}. The equations read (with vector indices running over radial, $r$, polar, $\theta$, and azimuthal, $\phi$, coordinate, and $\partial_\phi=0$):
\begin{subequations}\label{eq:hydevo}
\begin{align}
   \partial_t \rho + \nabla_j(\rho v^j) & = 0  \, , \\
   \partial_t (\rho Y_e) + \nabla_j(\rho Y_e v^j) &= Q_{\mathrm{N}}\, , \\
   \partial_t (\rho v^i) + \nabla_j(\rho v^i v^j + P_{\mathrm{g}} ) 
   & = - \rho\nabla^i\Phi +Q_{\mathrm{M}}^i  \, , \\   
   \partial_t e_{\mathrm{t}} + \nabla_j(v^j e_{\mathrm{t}} + v^j P_{\mathrm{g}}) \
   & = -\rho v_j\nabla^j\Phi + Q_{\mathrm{E}} + v_j Q_{\mathrm{M}}^j  \, , \label{eq:hydetotevo}
\end{align}
\end{subequations}
where $\rho, v^i, Y_e, e_{\mathrm{t}}, P_{\mathrm{g}}$, and $\Phi$ are the baryon density, fluid velocity, electron fraction, total (kinetic plus internal) gas energy density, gas pressure, and gravitational potential, respectively, and the neutrino-related source terms are given by (with neutrino-energy coordinate $\eps$ and baryonic mass constant $m_{\mathrm{B}}$)
\begin{subequations}\label{eq:qterms}
\begin{align}
 Q_{\mathrm{N}} &= - \alpha\, m_{\mathrm{B}}\int_0^\infty(S^{(0)}_{\nu_e}-S^{(0)}_{\bar\nu_e})\frac{\mathrm{d}\eps}{\eps} \, , \\
 Q_{\mathrm{M}}^i& =  - \frac{\alpha}{c^2}\sum_{\nu}\int_0^\infty S^{(1),i}_{\nu}\,\mathrm{d}\eps   \, , \\
 Q_{\mathrm{E}}& =  - \alpha\sum_{\nu}\int_0^\infty S^{(0)}_{\nu}\,\mathrm{d}\eps   \, ,
\end{align}
\end{subequations}
in terms of the lapse function $\alpha$ and the 0th- and 1st-order angular moments of the collision integral measured in the comoving (i.e. fluid rest) frame, $S^{(0)}_{\nu}$ and $S^{(1),i}_{\nu}$, respectively, where the index $\nu$ runs over all six neutrino species.

Both $\Phi$ and $\alpha$ are computed from radial profiles of angle averaged fluid- and neutrino-related quantities using the prescriptions of \citet{Marek2006} (for the case ``A'' potential) and \citet{Rampp2002}, respectively. We do not include spherical harmonics terms higher than the monopole in the gravitational potential, i.e. gravity is considered to be spherically symmetric.

The hydrodynamics module coupled to \textsc{Alcar}, \textsc{Aenus} \citep{Obergaulinger2008a}, offers several choices for spatial reconstruction methods, Riemann solvers, and time integration schemes. In this study, we employ spatial reconstruction as adapted from the piecewise-parabolic method \citep[PPM;][]{Colella1984} in the version by \citet{Mignone2014a} that retains high-order accuracy near the coordinate center and polar axis\footnote{More specifically, we use the cell-centered version ``PPM$_5$'' of \citet{Mignone2014a} to reconstruct the primitive variables and employ flattening near strong shocks as in \citet{Colella1984} but no steepening near contact discontinuities. We do not perform characteristic tracing as in the original PPM method.}, the HLLC Riemann solver everywhere except near coordinate-aligned shocks where we switch to the HLLE solver \citep[e.g.][]{Toro1997}, and a dimensionally unsplit 2nd-order Runge-Kutta time-integration scheme.

The neutrino transport in both codes is based on the multi-group evolution of the specific (i.e. per unit of neutrino energy) energy density, $E$, and specific flux density, $F^i$, both measured in the comoving frame of the fluid, for the three species electron-neutrinos $\nu_e$, electron-antineutrinos $\bar\nu_e$, as well as $\nu_x$ representative of the four remaining heavy-lepton neutrinos. The latter is governed by (suppressing indices $\nu$):
\begin{subequations}\label{eq:momeq}
\begin{align}
  &\partial_tE + \nabla_j\left(\alpha F^j + v^j E\right) +
  P^{ij}\nabla_iv_j + F^i \nabla_i\alpha \nonumber\\
  &\hspace{1.5cm}- \partial_\eps\left[\eps\left(  P^{ij} \nabla_iv_j + 
      F^i\nabla_i\alpha\right)\right] = \alpha S^{(0)} \, , \\
  &\partial_tF^i + \nabla_j\left(\alpha c^2 P^{ij} + v^j F^i\right) +
  F^j\nabla_jv^i + c^2 E \nabla^i\alpha \nonumber\\
  &\hspace{1.5cm}-\partial_\eps\left[\eps\left(Q^{ijk}\nabla_j v_k + 
      c^2 P^{ij} \nabla_j\alpha \right)\right] = \alpha S^{(1),i} \, ,
\end{align}
\end{subequations}
where $P^{ij}, Q^{ijk}$ are the 2nd- and 3rd-order angular moment tensors, respectively. Compared to the purely Newtonian counterparts, Eqs.~(\ref{eq:momeq}) contain corrections for general relativistic redshift and time dilation. We use the same corrections in \textsc{Alcar} as those in \textsc{Vertex}, which have been motivated in \citet[][]{Rampp2002}\footnote{Specifically, equations~(\ref{eq:momeq}) can be recovered from the general relativistic moment equations \citep[e.g.][where necessary after transforming the moments into the comoving frame]{Shibata2011, Cardall2013a} by assuming the only general relativistic metric component to be the lapse function, $\alpha$, performing the replacement $\alpha v^i\rightarrow v^i$, and dropping all terms proportional to $\mathcal{O}(v^2/c^2)$ and $\mathcal{O}(v/c\nabla\alpha)$.} and shown, e.g. in \citet{Liebendorfer2005, Muller2012b}, to lead to results that agree reasonably well with fully general relativistic results.

In \textsc{Alcar}, the  higher-order moments $P^{ij}$ and $Q^{ijk}$ needed to close the set of moment equations are expressed as functions of the moments $E, F^i$ by
\begin{subequations}\label{eq:eddtensor}
\begin{align}
  \frac{P^{ij}}{E} & = \frac{1-\chi}{2}\,\delta^{ij} + \frac{3\chi - 1}{2}\, n^i_\textbf{\em F}\,n^j_\textbf{\em F} \, ,\\
  \frac{Q^{ijk}}{E} &=   \frac{f-q}{2} (n^i_\vecF\,\delta^{jk} \!+ \!
    n^j_\vecF\,\delta^{ik} \!+ \! n^k_\vecF\,\delta^{ij} ) \! + \!\frac{5q-3f}{2}\, n^i_\vecF \,n^j_\vecF\, n^k_\vecF \, ,
\end{align}
\end{subequations}
where $f\equiv |\vecF|/(c E)$, $n^i_\vecF\equiv F^i/|\vecF|$, and the quantities $\chi$ and $q$ are determined by the chosen one-dimensional closure. For the latter, we use the Minerbo-closure and refer to \citet{Minerbo1978} and \citet{Just2015b} for explicit expressions of $\chi$ and $q$, respectively. The resulting hyperbolic system of equations is solved in close analogy to the hydrodynamics equations, namely on the same grid and with the same reconstruction scheme, and an HLL Riemann solver. The time integration of the potentially stiff interaction source terms, however, is done in a mixed explicit-implicit manner in order to ensure numerical stability (see Appendix~\ref{sec:comp-source-terms} for details).

\textsc{Vertex} solves the same set of equations, Eqs.~(\ref{eq:hydevo}) and (\ref{eq:momeq}), for hydrodynamics and neutrino transport, respectively, except for the following differences: \textsc{Vertex} obtains the higher-order moments using a variable-Eddington-factor approach, i.e. by solving a (slightly simplified) Boltzmann equation in addition to the moment equations. This  approach is comparable in accuracy to solving the Boltzmann equation directly for a spherically symmetric stellar background and is then constrained to be used in the RbR+ mode. The latter assumes that the radiation field is axisymmetric around each radial ray, leading to vanishing non-radial components of the flux density vector, $F^i$, while lateral advection of neutrinos by the fluid as well as lateral neutrino-pressure forces are still taken into account \citep[see][for details]{Buras2006}. \textsc{Vertex} evolves Eqs.~(\ref{eq:momeq}) using finite-difference methods and fully implicit time integration. For the hydrodynamics part the well-known \textsc{Prometheus} code \citep{Fryxell1989a} is applied, which employs the original PPM method \citep{Colella1984} to integrate the hydrodynamics equations with 2nd-order Strang-type dimensional splitting.

\subsection{Model setup and neutrino interaction channels}\label{sec:model-setup}

We consider two stellar progenitor models in this study, one with rather high zero-age main sequence (ZAMS) mass of $20\,\Msol$, model s20 \citep{Woosley2007a}, and one with lower ZAMS mass of $9\,\Msol$ (model 9.0A of \citealp{Woosley2015a}\footnote{We use the slightly upgraded version with respect to \citet{Woosley2015a} that was also employed by \citealp{Sukhbold2016a} and that is available at \url{https://wwwmpa.mpa-garching.mpg.de/ccsnarchive/data/SEWBJ_2015/index.html}.}), which we denote here as model s9. The simulations are initialized using the density, temperature, and electron fraction from the progenitor data. The 2D models are set up by mapping from 1D simulations at around $15-20$\,ms after bounce and adding random perturbations $\delta\rho \in [-10^{-3}\rho,10^{-3}\rho]$ to the density, $\rho$, in each grid cell. In doing so, we ignore recent findings \citep[e.g.][]{Couch2015a,Muller2016c} that rather strong perturbations with low angular order can be present before collapse, which may help initiating the shock runaway. Since in this study we are only concerned about the impact of modeling variations on the post-bounce evolution, all two-dimensional models are mapped from the same 1D reference model corresponding to the code and progenitor (e.g. s20-pp-str-mb-norel is mapped from s20-ref-1D; cf. Table~\ref{table_models} and Sect.~\ref{sec:neutr-inter-invest}) -- hence the collapse and bounce are always simulated without any transport simplifications. We note, however, that focussing here only on the post-bounce evolution does not mean that any of the considered modeling variations is insignificant when applied during the collapse phase. In fact, some variations may even lead to more dramatic consequences than observed here when used during the collapse \citep[e.g.][]{Arnett1977a, Bruenn1985, Lentz2012a}.

We employ the equation of state (EOS) ``SFHo'' of \cite{Steiner2013}. Since this EOS is constructed assuming nuclear statistical equilibrium (NSE), ideally we should replace it by a composition-dependent EOS linked to a nuclear reaction network once the temperature drops below $\sim 5\,$GK. However, for the present study we avoid this additional level of complexity and employ the SFHo-table everywhere. Any neutrino interactions with light nuclear clusters (i.e. elements with mass numbers $A=2-4$ except $\alpha$ particles) appearing in the SFHo EOS are ignored. The original SFHo table only allows electron fractions up to $0.6$. This turned out to be insufficient for a subset of models in which transient, proton-rich bubbles arise in the gain region. To this end, we extended the original SFHo table in the sub-nuclear domain towards more proton-rich conditions using a 23-species NSE solver.

For the \textsc{Alcar} simulations, the radial grid remains fixed during the simulations. The standard resolution has $N_r=640$ radial zones with a width of $\Delta r=300\,$m below $r=30\,$km and growing by a constant factor per cell up to $\Delta r=2\,$km at $r=300\,$km and finally increasing by $\approx 2\,\%$ per cell up to the outer grid boundary at $r=10^9\,$cm. In the \textsc{Vertex} simulations the radial grid at the time of bounce consists of $400$ zones, the distribution of which results from a quasi-Lagrangian treatment of the collapse, while during the post-bounce evolution the radial grid remains fixed (i.e. Eulerian) but is manually refined multiple times around the neutrinosphere by using a density-based criterion. Typically, a final number of radial zones of $N_r\sim 570-630$ is reached at the end of the simulations. The angular grids are always uniform in $\theta$, with the standard number of zones being $N_\theta=240$ and $160$ for \textsc{Alcar} and \textsc{Vertex}, respectively. The neutrino energy space is discretized using 15 energy groups distributed nearly logarithmically between 0 and $400\,$MeV (\textsc{Alcar}) or $380\,$MeV (\textsc{Vertex}).

With a spherical polar coordinate mesh the Courant time step would become prohibitively small near the coordinate center. We mitigate this problem by using a spherically symmetric core up to a certain radius. In \textsc{Alcar}, this radius decreases linearly in time from 10\,km at 20\,ms post bounce down to 7\,km at about 0.5\,s post bounce, whereafter it remains constant. In \textsc{Vertex}, the radius of the 1D core is 1.6\,km at all times.

Our reference models include the following set of neutrino reactions (and corresponding reverse reactions):
\begin{itemize}[leftmargin=3mm]
\item Electron/positron captures by protons/neutrons following \citet{Bruenn1985} including the corrections by \citet{Horowitz2002} for weak magnetism and nucleon recoil.
\item Isoenergetic scattering of all neutrino types on neutrons and protons following \citet{Bruenn1985} and likewise augmented with corrections for weak magnetism and nucleon recoil \citep{Horowitz2002}.
\item Electron captures by heavy nuclei as in \citet{Bruenn1985}.
\item Coherent scattering of all neutrino types off heavy nuclei including corrections due to the nuclear form factor and ion-ion correlations \citep[cf.][]{Bruenn1997, Horowitz1997a}.
\item Inelastic scattering of all neutrino types off electrons and positrons \citep{Yueh1977, Bruenn1985,Cernohorsky1994a}.
\item Pair processes, namely pair production and annihilation of electron-type (i.e. $\nu_e$ and $\bar\nu_e$) as well as heavy-lepton neutrinos via nucleon-nucleon bremsstrahlung and electron-positron annihilation and pair production as in \citet{Hannestad1998} and \citet{Pons1998}, respectively.
\end{itemize}
Motivated by a number of recent studies \citep{Melson2015a,Bollig2017a,Radice2017a,OConnor2017a} we additionally include the following corrections in some models:
\begin{itemize}[leftmargin=3mm]
\item Strangeness correction to the axial-vector coupling in neutral-current neutrino-nucleon interactions with $g_a^s=-0.1$ \citep{Horowitz2002, Horowitz2017a}.
\item Axial response corrections to the neutrino-nucleon scattering cross section due to many-body effects as in \citep[][using the fit formula for $S^f_A$]{Horowitz2017a}. 
\end{itemize}
However, the impact of these opacity upgrades on the post-bounce evolution has been discussed before, so the main reasons for including them here are, first, to provide a reference for comparison with the considered modeling simplifications, and second, to enhance for some models the explodability in order to test modeling simplifications in an extended range of conditions (see Sect.~\ref{sec:neutr-inter-invest} for the investigated list of models).

For the implementation of the rates in \textsc{Alcar} we followed as closely as possible that of \textsc{Vertex}, which is described in the Appendix of \citet{Rampp2002}. However, all neutrino rates (as well as all other physics modules) in \textsc{Alcar} have been coded from scratch, i.e. no routines have been copied from \textsc{Vertex}. The numerical treatment of the source terms in \textsc{Alcar} differs from that of \textsc{Vertex} in that it, first, allows for three instead of only one flux-vector component, and second, is both explicit and implicit in time depending on the type of interaction and dynamic conditions, whereas it is fully implicit in \textsc{Vertex}. We provide more details on the implementation of the source terms in \textsc{Alcar} in Appendix~\ref{sec:comp-source-terms}.

We note that the set of neutrino interactions chosen here is not as advanced as the one usually employed in \textsc{Vertex} \citep[e.g.][]{Buras2006}. However, we chose this rather basic set to facilitate the comparison between the two codes and to allow other groups to compare with our results on the basis of a widely available repository of input physics.

\setlength{\tabcolsep}{2.3pt}
\begin{table*}
  \caption{Overview of models considered in this study and their properties. The columns contain from left to right: Model name, employed simulation code, progenitor model, number of radial ($N_r$) and angular ($N_\theta$) grid zones (where $N_\theta=1$ means that spherical symmetry was assumed), information about the inclusion of neutrino-electron scattering, treatment of pair processes, inclusion of the strangeness correction (with $g_a^s=-0.1$), inclusion of many-body corrections (using the fit for $S^f_A$ of \citealp{Horowitz2017a}), inclusion of velocity-dependent and (Doppler and gravitational) energy-shift terms in the transport, the assumption of RbR+, and for exploding models the explosion time defined here as the post-bounce time when the angle-averaged shock surface reaches $r=300$\,km. Each slash-separated number at the end of a model name labels an additional simulation initialized with a different random perturbation pattern, for which the runaway times are correspondingly given, separated by slashes.}
  \label{table_models}
  \begin{center}
  \begin{tabularx}{\textwidth}{lcccccccccl}
    \hline \hline
    model                       & simulation      & prog. & $N_r\times N_\theta$       & $\nu$-$e^\pm$ & pair-proc. & strangeness & many-body  & $\frac{v}{c}$ and $\partial_\eps$ & RbR+ & explosion      \\
    name                        & code            & model &                            & scatt.        & treatment  & correction  & correction & terms                             &      & time [s]       \\
    \hline                                                                                                                           
    s20-ref-1D                  & \textsc{Alcar}  & s20   & $640\times 1$              & yes           & full       & no          & no         & yes                               & --   & --             \\
    s20-nones-1D                & \textsc{Alcar}  & s20   & $640\times 1$              & no            & full       & no          & no         & yes                               & --   & --             \\
    s20-pp-1D                   & \textsc{Alcar}  & s20   & $640\times 1$              & yes           & simple     & no          & no         & yes                               & --   & --             \\
    s20-str-1D                  & \textsc{Alcar}  & s20   & $640\times 1$              & yes           & full       & yes         & no         & yes                               & --   & --             \\
    s20-mb-1D                   & \textsc{Alcar}  & s20   & $640\times 1$              & yes           & full       & no          & yes        & yes                               & --   & --             \\
    s20-norel-1D                & \textsc{Alcar}  & s20   & $640\times 1$              & yes           & full       & no          & no         & no                                & --   & --             \\
    \hline                                                                                                                           
    s20-ref\{1/2/3\}            & \textsc{Alcar}  & s20   & $640\times 240$            & yes           & full       & no          & no         & yes                               & no   & --/--/--       \\
    s20-str-mb\{1/2\}           & \textsc{Alcar}  & s20   & $640\times 240$            & yes           & full       & yes         & yes        & yes                               & no   & --/--          \\
    s20-pp\{1/2\}               & \textsc{Alcar}  & s20   & $640\times 240$            & yes           & simple     & no          & no         & yes                               & no   & 1.14/1.10      \\
    s20-pp-str\{1/2/3\}         & \textsc{Alcar}  & s20   & $640\times 240$            & yes           & simple     & yes         & no         & yes                               & no   & 0.98/0.80/0.67 \\
    s20-pp-mb                   & \textsc{Alcar}  & s20   & $640\times 240$            & yes           & simple     & no          & yes        & yes                               & no   & 1.03           \\
    s20-pp-str-mb\{1/2/3\}      & \textsc{Alcar}  & s20   & $640\times 240$            & yes           & simple     & yes         & yes        & yes                               & no   & 0.41/0.50/0.38 \\
    s20-pp-str-mb-norel         & \textsc{Alcar}  & s20   & $640\times 240$            & yes           & simple     & yes         & yes        & no                                & no   & 0.31           \\
    s20-rbr\{1/2/3\}            & \textsc{Alcar}  & s20   & $640\times 240$            & yes           & full       & no          & no         & yes                               & yes  & 0.81/0.48/0.92 \\
    s20-rbr-norel               & \textsc{Alcar}  & s20   & $640\times 240$            & yes           & full       & no          & no         & no                                & yes  & --             \\
    s20-rbr-nones               & \textsc{Alcar}  & s20   & $640\times 240$            & no            & full       & no          & no         & yes                               & yes  & --             \\
    s20-rbr-pp\{1/2/3\}         & \textsc{Alcar}  & s20   & $640\times 240$            & yes           & simple     & no          & no         & yes                               & yes  & 0.33/0.36/0.37 \\
    s20-rbr-pp-nones            & \textsc{Alcar}  & s20   & $640\times 240$            & no            & simple     & no          & no         & yes                               & yes  & 0.38           \\
    s20-ref-hires               & \textsc{Alcar}  & s20   & $960\times 320$            & yes           & full       & no          & no         & yes                               & no   & --             \\
    s20-ref-lores               & \textsc{Alcar}  & s20   & $320\times 120$            & yes           & full       & no          & no         & yes                               & no   & --             \\
    s20-rbr-hires               & \textsc{Alcar}  & s20   & $960\times 320$            & yes           & full       & no          & no         & yes                               & yes  & 0.59           \\
    s20-rbr-lores               & \textsc{Alcar}  & s20   & $320\times 120$            & yes           & full       & no          & no         & yes                               & yes  & 0.85           \\
    s20-rbr-hi$\theta$\{1/2/3\} & \textsc{Alcar}  & s20   & $640\times 320$            & yes           & full       & no          & no         & yes                               & yes  & 0.68/0.43/0.86 \\
    s20-rbr-lo$\theta$\{1/2/3\} & \textsc{Alcar}  & s20   & $640\times 80$             & yes           & full       & no          & no         & yes                               & yes  & 0.64/0.44/0.77 \\
    \hline                                                                                                                           
    s20VX-1D                    & \textsc{Vertex} & s20   & $\sim$(400-600)$\times 1$  & yes           & full       & no          & no         & yes                               & --   & --             \\
    s20VX-nones-1D              & \textsc{Vertex} & s20   & $\sim$(400-600)$\times 1$  & no            & full       & no          & no         & yes                               & --   & --             \\
    \hline                                                                                                                           
    s20VX\{1/2\}                & \textsc{Vertex} & s20   & $\sim$(400-600)$\times$160 & yes           & full       & no          & no         & yes                               & yes  & 0.53/0.70      \\
    s20VX-nones\{1/2\}          & \textsc{Vertex} & s20   & $\sim$(400-600)$\times$160 & no            & full       & no          & no         & yes                               & yes  & --/--          \\
    \hline                                                                                                                           
    s9-ref-1D                   & \textsc{Alcar}  & s9    & $640\times 1$              & yes           & full       & no          & no         & yes                               & --   & --             \\
    s9-nones-1D                 & \textsc{Alcar}  & s9    & $640\times 1$              & no            & full       & no          & no         & yes                               & --   & --             \\
    s9-pp-1D                    & \textsc{Alcar}  & s9    & $640\times 1$              & yes           & full       & no          & no         & yes                               & --   & --             \\
    s9-str-mb-1D                & \textsc{Alcar}  & s9    & $640\times 1$              & yes           & full       & yes         & yes        & yes                               & --   & --             \\
    s9-norel-1D                 & \textsc{Alcar}  & s9    & $640\times 1$              & yes           & full       & no          & no         & no                                & --   & --             \\
    \hline                                                                                                                           
    s9-ref\{1/2/3\}             & \textsc{Alcar}  & s9    & $640\times 240$            & yes           & full       & no          & no         & yes                               & no   & 0.41/0.45/0.44 \\
    s9-nones\{1/2\}             & \textsc{Alcar}  & s9    & $640\times 240$            & no            & full       & no          & no         & yes                               & no   & --/--          \\
    s9-rbr\{1/2/3\}             & \textsc{Alcar}  & s9    & $640\times 240$            & yes           & full       & no          & no         & yes                               & yes  & 0.35/0.39/0.32 \\
    s9-rbr-nones                & \textsc{Alcar}  & s9    & $640\times 240$            & no            & full       & no          & no         & yes                               & yes  & --             \\
    s9-pp                       & \textsc{Alcar}  & s9    & $640\times 240$            & yes           & simple     & no          & no         & yes                               & no   & 0.32           \\
    s9-str-mb                   & \textsc{Alcar}  & s9    & $640\times 240$            & yes           & full       & yes         & yes        & yes                               & no   & 0.31           \\
    s9-norel                    & \textsc{Alcar}  & s9    & $640\times 240$            & yes           & full       & no          & no         & no                                & no   & 0.41           \\
    \hline \hline
  \end{tabularx}
\end{center}
\end{table*}

\subsection{Investigated models}\label{sec:neutr-inter-invest}


The axisymmetric reference models run with \textsc{Alcar} are s20-ref and s9-ref for the s20 and s9 progenitors, respectively, and evolve the fully multidimensional (i.e. not RbR-constrained) transport equations, Eqs.~\eqref{eq:momeq}, using the aforementioned neutrino interactions (except strangeness and many-body corrections). Models with ``str'' and ``mb'' in their names additionally include the strangeness and many-body corrections, respectively. Models with ``rbr'' in their names adopt the RbR+ approximation in \textsc{Alcar} exactly as described in \citet{Buras2006}, namely by setting the lateral flux densities\footnote{We note that this also includes the diffusive component of the HLL fluxes through the cell interfaces, which can be non-zero even if the cell-centered fluxes vanish.}, $F_\theta=F_\phi=0$ (though $F_\phi=0$ is fulfilled anyway in our non-rotating models) and setting the lateral source term for the evolution of $\rho v^\theta$,  $Q_{\mathrm{M}}^\theta$, to
\begin{equation}
  Q_{\mathrm{M}}^{\theta,\mathrm{RbR}} = -\frac{\alpha}{3}\sum_{\nu}\int_\eps \frac{1}{r\sin\theta}\partial_\theta E_{\nu}\,\mathrm{d}\eps 
\end{equation}
for trapping densities, $\rho > 10^{12}$g\,cm$^{-3}$, and to zero for lower densities. In this way, lateral advection and compression of energy and radial flux remain included in the transport equations as well as lateral neutrino-pressure forces in the gas-momentum equations.

The remaining modeling variations are: Turning off inelastic neutrino-electron scattering (``nones''), ignoring velocity-dependent and gravitational redshift terms (``norel'') by setting $v=0$ and $\partial_\eps=0$ in Eqs.~\eqref{eq:momeq}, and using the simplified description of ($e^\pm$-annihilation and bremsstrahlung) pair processes (``pp'') as suggested in \citet{OConnor2015a}. The pair-process simplification consists of, first, ignoring all pair processes for electron-type neutrinos entirely, and second, assuming that respective annihilation partners for pair-annihilation of $\nu_x$ are in isotropic, local thermodynamic equilibrium (LTE) when computing the pair-process interaction kernel. The second assumption reduces the source terms for $\nu_x$ pair processes to be formally equivalent to source terms for emission/absorption processes \citep[see][for more details]{OConnor2015a}. All three aforementioned simplifications are particularly appealing for energy-dependent transport schemes, because each of them entails dropping numerically complicated energy-bin coupling terms. We include these simplifications here, because they have rarely been tested so far in multidimensional simulations\footnote{One exception is inelastic neutrino-electron scattering, which was found in \citet{Burrows2018a} to have a similar impact as seen in this paper. For 1D models the impact of neglecting neutrino-electron scattering and velocity terms has been tested in \citet{Lentz2012a,Lentz2012b}. However, those studies did not test the impact just on the post-bounce evolution (i.e. starting from the same post-bounce models), which is what we do here.}.

The list of all simulations is provided in Table~\ref{table_models}. The two \textsc{Alcar} models using RbR+ with and without neutrino-electron scattering (s20-rbr and s20-rbr-nones, respectively) are compared with the two \textsc{Vertex} models s20VX and s20VX-nones that contain exactly the same corresponding input physics. The setup of the remaining models is motivated mainly to test the impact of each modeling simplification (represented by ``rbr'', ``nones'', ``norel'', ``pp'') on the eventual onset time of explosion, and to compare this impact with that of using the opacity improvements labeled by ``str'' and ``mb''. For the s20 model\footnote{The reader might wonder why for some of the one-dimensional s20 models the corresponding 2D counterparts are missing. The reason simply is that these counterparts are too unlikely to explode given the existing results, and therefore were deemed less informative for this study.}, each modeling simplification is tested for at least two models with different conditions regarding the proximity to explosion; e.g., we turn off neutrino-electron scattering both for rather late (s20-rbr/-nones) and early (s20-rbr-pp/-nones) exploding models. Since the reference s20 model, s20-ref, does not lead to an explosion, we use in some cases its exploding RbR+ counterpart, s20-rbr, as reference for the comparison. Additional models (s20-pp-str, s20-pp-mb, and s20-pp-str-mb) are set up to test which and how many opacity variations are needed to push the reference s20 model to an early explosion.

In order to obtain a rough idea about the influence of stochasticity, we perform for some models additional simulations that differ only in the pattern (but not amplitude) of initial random density perturbations. These simulations are labeled by numbers at the end of the model names. When discussing these models below we will suppress these numbers if the point of concern holds for all simulations of the given model independently of the initial perturbation pattern. Finally, several simulations with different numbers of grid cells are conducted in order to test the resolution dependence.

All simulations are stopped either once shock expansion sets in or after 1\,s (0.9\,s) of post-bounce evolution in \textsc{Alcar} (\textsc{Vertex}), except for individual models that are followed slightly longer because of optimistic runaway conditions.

\section{Results: 1D models}\label{sec:results:-1d-models}

\begin{figure*}
  \centering
  \includegraphics[width=\textwidth]{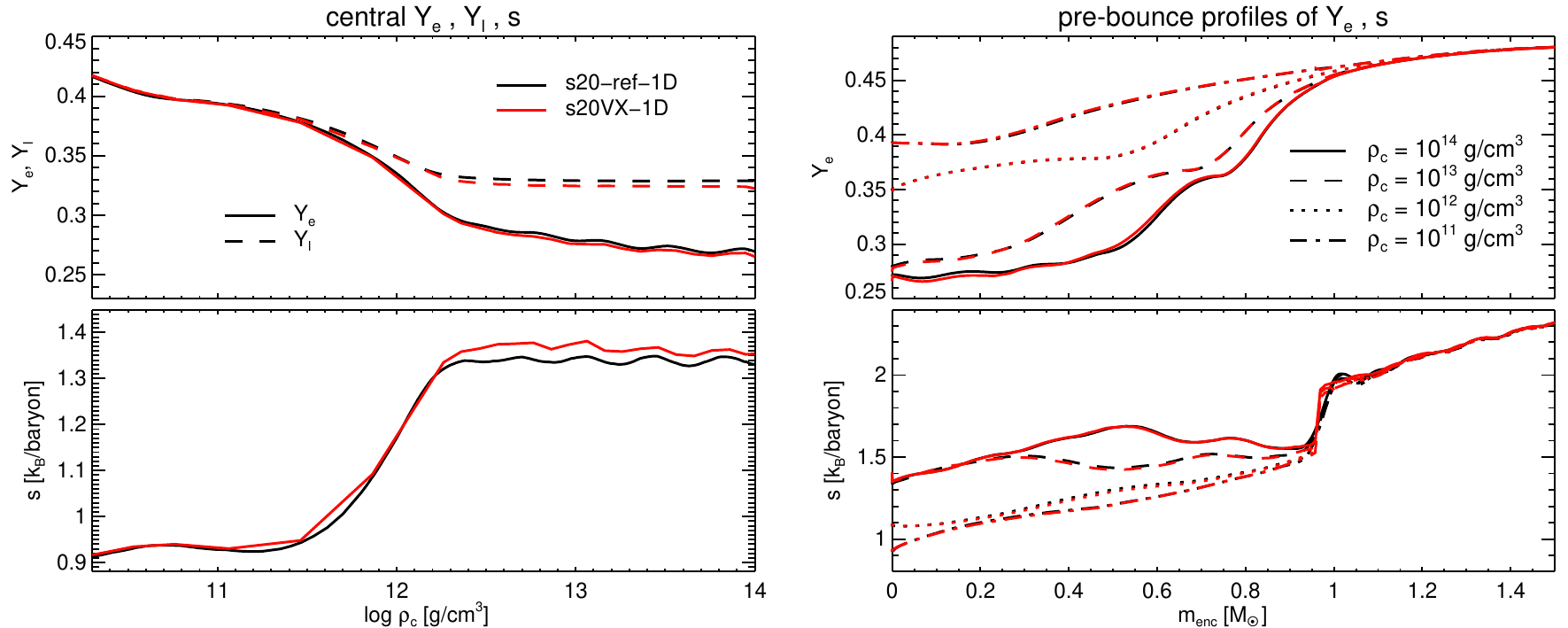}
  \caption{Comparison between \textsc{Alcar} and \textsc{Vertex} models of properties characterizing the collapse. The left panel shows the central electron ($Y_e$) and lepton ($Y_l$) fraction and central entropy per baryon ($s$) as functions of central density. The right panel shows profiles of $Y_e$ and $s$ with respect to the enclosed mass for times at which the central density reaches the indicated values.}
  \label{fig:collapse}
\end{figure*}	

\begin{figure*}
  \centering 
  \includegraphics[width=0.49\textwidth]{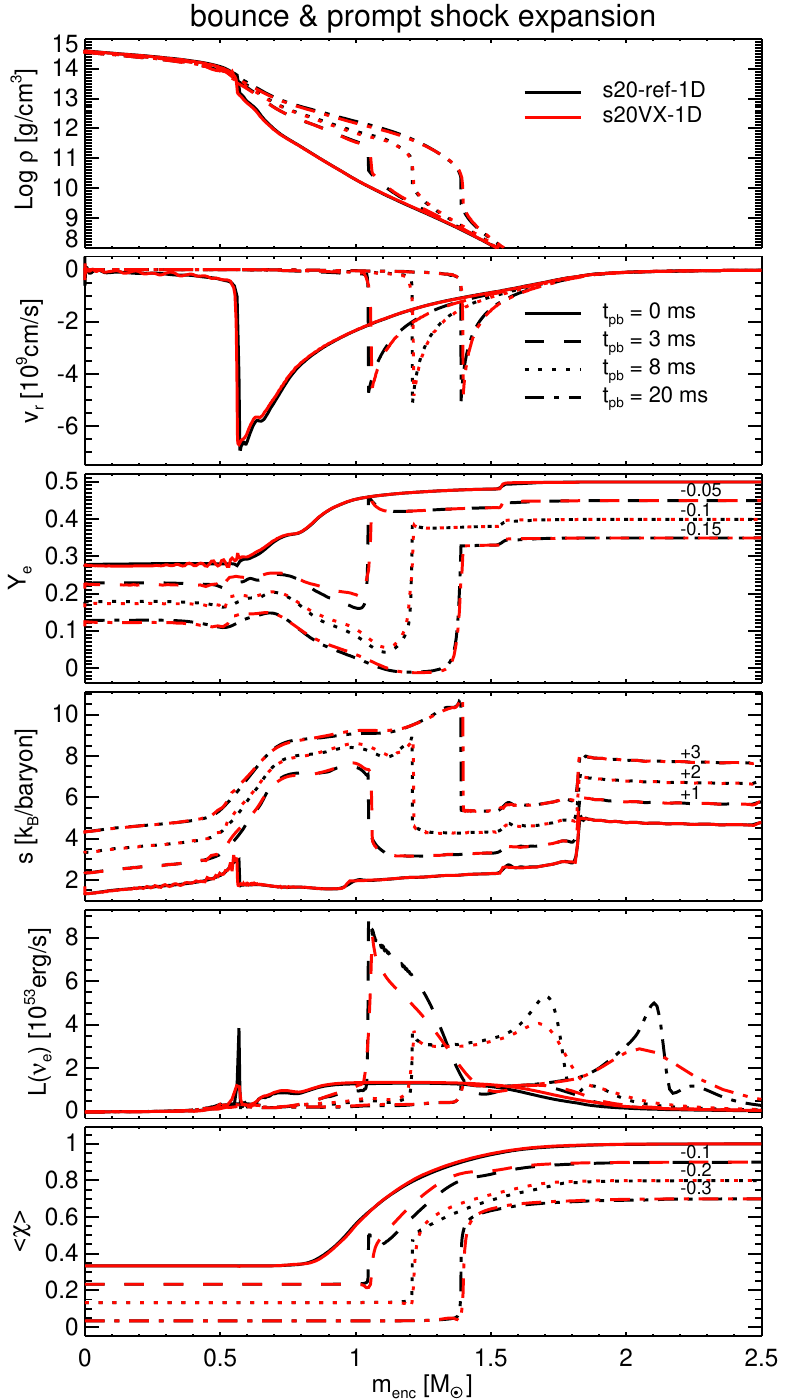}
  \includegraphics[width=0.49\textwidth]{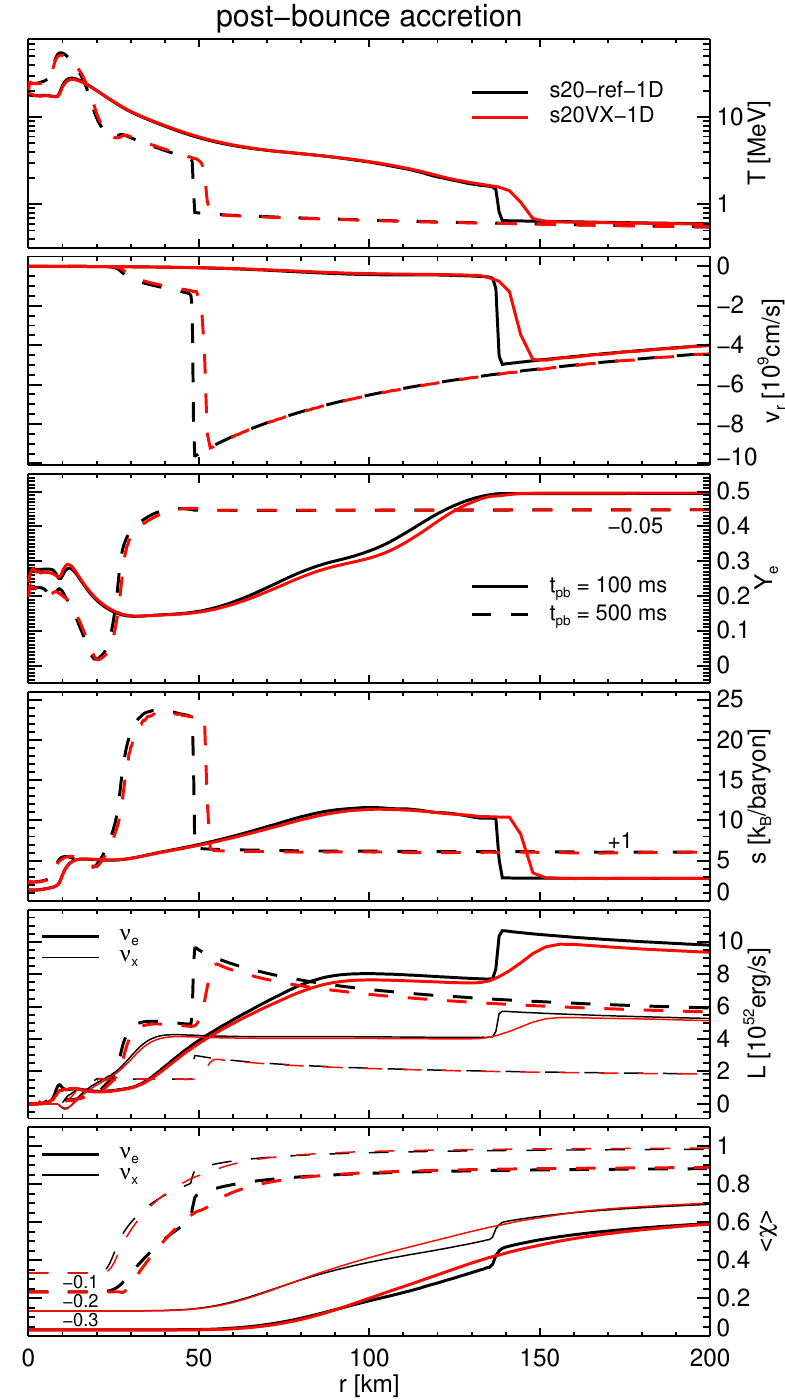}
  \caption{Profiles of hydrodynamic and neutrino-related quantities as functions of enclosed mass (left) and radius (right) for two spherically symmetric \textsc{Alcar} and \textsc{Vertex} models at the indicated times. The left panels provide from top to bottom the density, radial velocity, electron fraction, entropy, and the luminosity as well as mean Eddington factor of electron neutrinos computed via Eqs.~\eqref{eq:lumi} and \eqref{eq:eddfac}, respectively, using comoving-frame quantities. The right panels show the same quantities except that the density is replaced by the temperature. Numbers associated with lines denote the offset in units of the corresponding $y$-axis by which the lines have been shifted.}
  \label{fig:profiles}
\end{figure*}	

\begin{figure}
  \centering 
  \includegraphics[width=\columnwidth]{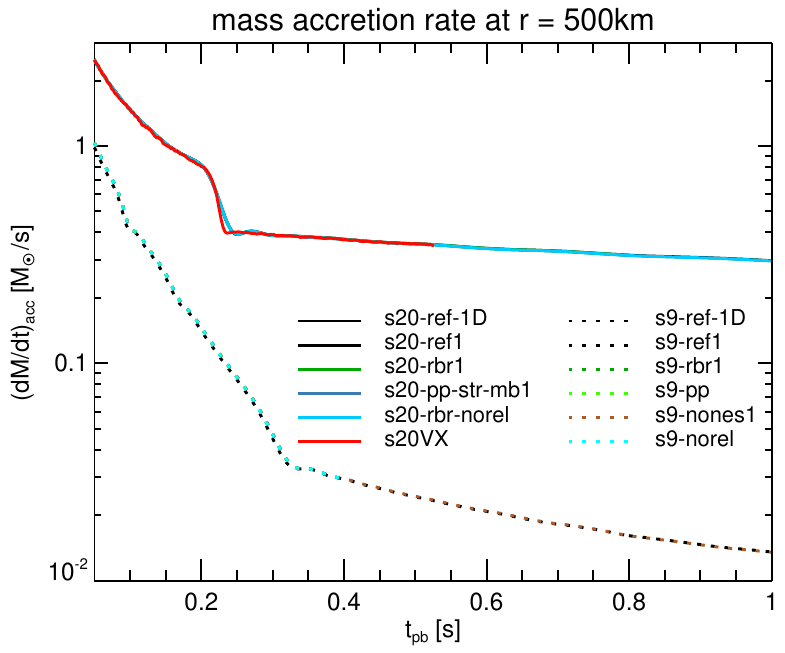}
  \caption{Comparison of mass accretion rates measured at a radius of $500$\,km for most of our models. Solid (dotted) lines refer to models using the s20 (s9) progenitor.}
  \label{fig:macc}
\end{figure}

\begin{figure*}
\centering
 \includegraphics[width=\textwidth]{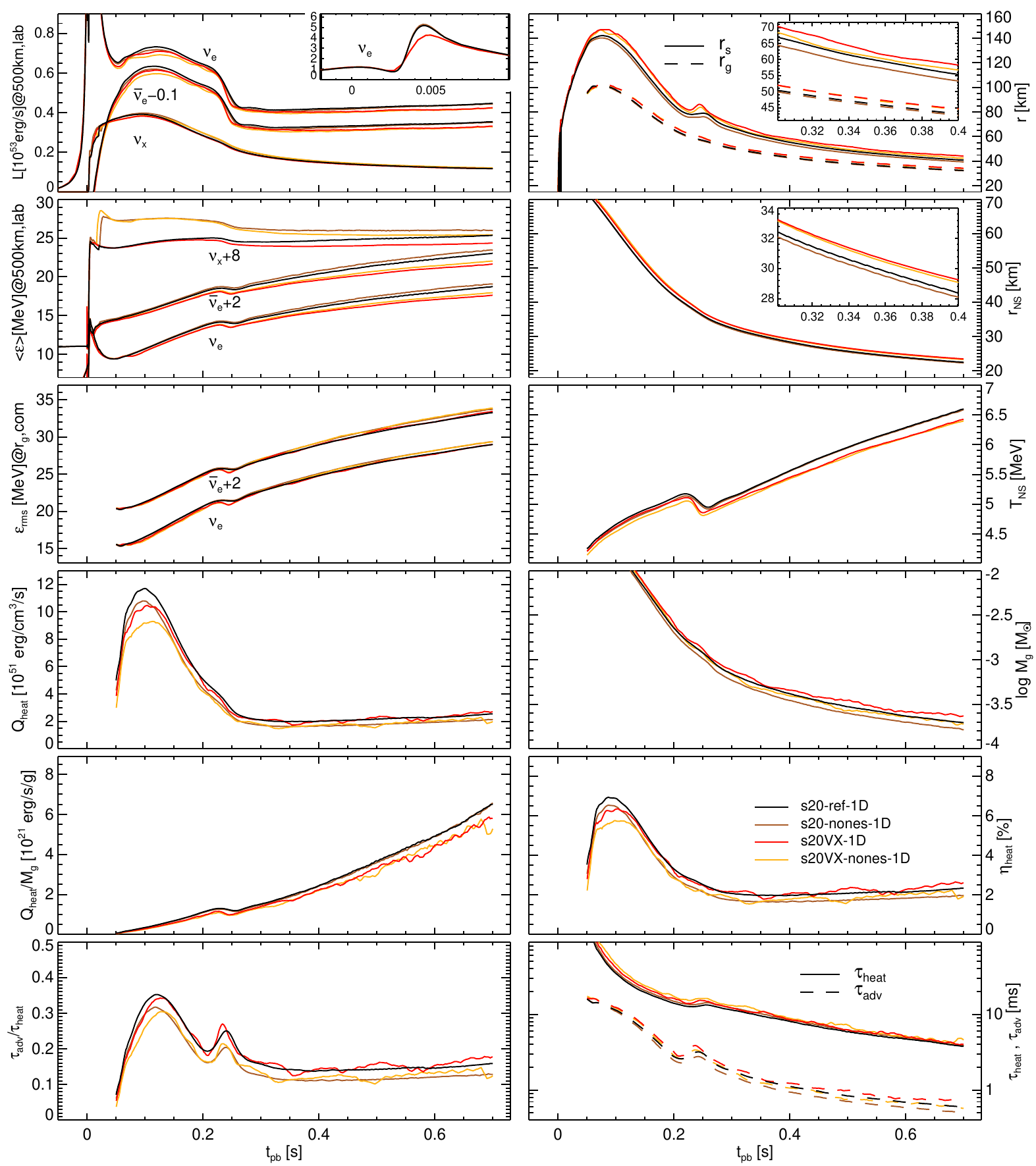}
 \caption{Comparison of global properties as functions of time for the spherically symmetric \textsc{Alcar} and \textsc{Vertex} models of the s20 progenitor whose names are displayed in the panel showing $\eta_{\mathrm{heat}}$. Left panels show from top to bottom the luminosities, $L$, and mean energies, $\langle\eps\rangle$, both measured in the lab-frame by an observer at infinity, the rms-energies, $\eps_{\mathrm{rms}}$, measured at the gain radius in the local comoving frame, the total and specific neutrino heating rate in the gain region, $Q_{\mathrm{heat}}$ and $Q_{\mathrm{heat}}/M_{\mathrm{g}}$, respectively, and the ratio of advection timescale to neutrino heating timescale, $\tau_{\mathrm{adv}}/\tau_{\mathrm{heat}}$. Right panels show from top to bottom the shock (gain) radius, $r_{\mathrm{s}}$ ($r_{\mathrm{g}}$), the radius, $r_{\mathrm{NS}}$, and temperature, $T_{\mathrm{NS}}$, of the PNS surface where $\rho=10^{11}\,$g\,cm$^{-3}$, mass of the gain region, $M_{\mathrm{g}}$, heating efficiency, $\eta_{\mathrm{heat}}$ (computed using Eq.~\eqref{eq:eta} with the lab-frame luminosities given in this plot), and the heating (advection) timescales, $\tau_{\mathrm{heat}}$ ($\tau_{\mathrm{adv}}$). Insets show enlarged regions of the same plots. Labels including numbers denote the margin in units of the current $y$-axis by which the curves for a given neutrino species are shifted, e.g. curves labeled $\bar\nu_e-0.1$ in the top left panel show the luminosity of electron anti-neutrinos reduced by $0.1\times 10^{53}\,$erg\,s$^{-1}$. All radii and $T_{\mathrm{NS}}$ show angle-averaged data. The curves are smoothed using running averages of 10\,ms.}
\label{fig:timplot1d1}
\end{figure*}	

\begin{figure*}
\centering
 \includegraphics[width=\textwidth]{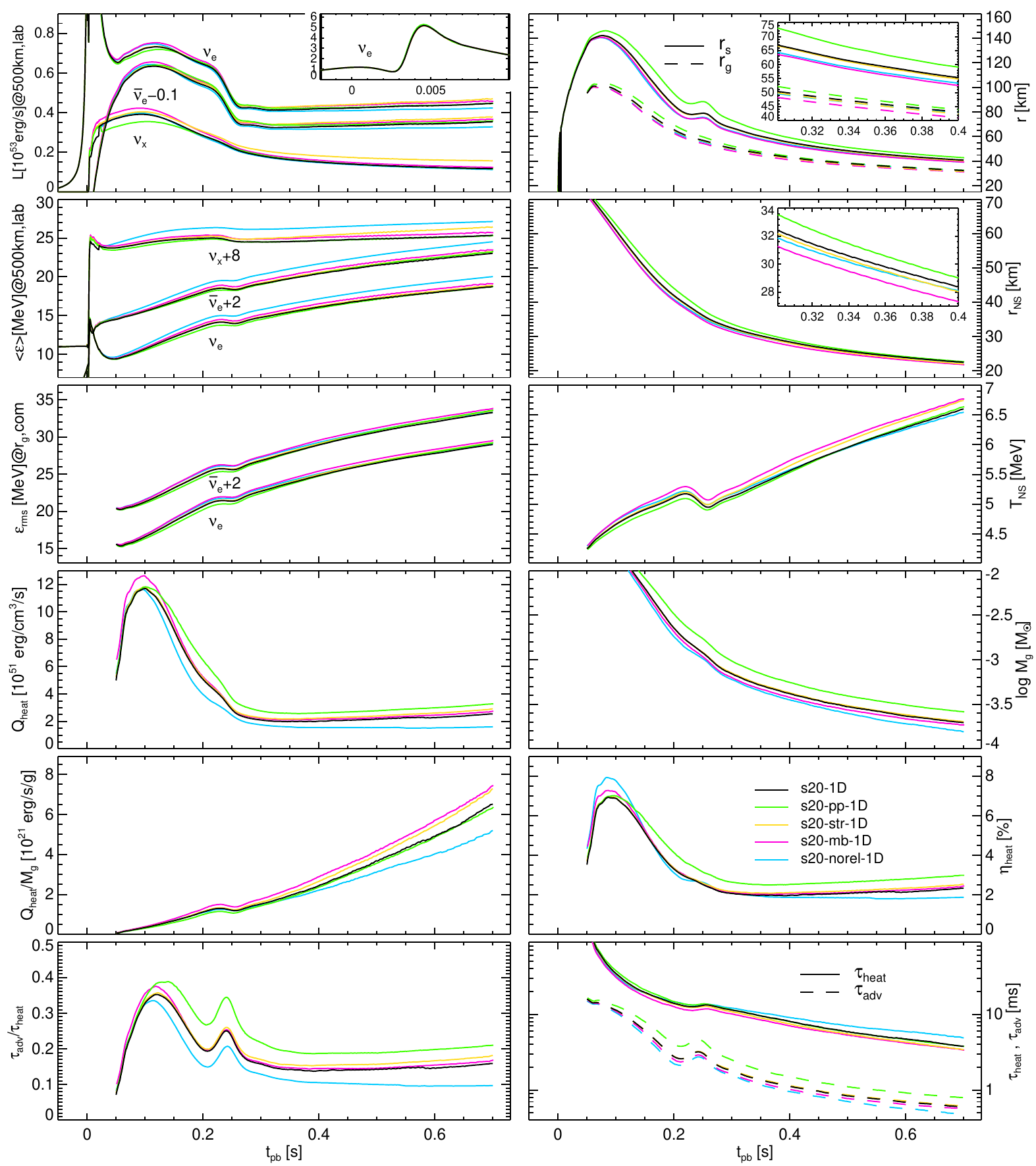}
 \caption{Same as Fig.~\ref{fig:timplot1d1} but for the remaining spherically symmetric \textsc{Alcar} models of the s20 progenitor.}
\label{fig:timplot1d2}
\end{figure*}	

We begin by comparing spherically symmetric models. Although rather unspectacular, 1D models
offer the most straightforward and computationally economic way to identify the basic impact of modeling variations, and they may reveal features that could remain obscured by stochasticity in multidimensional models. Since for non-exploding 1D models the progenitor dependence is less relevant, we only discuss the s20 models in this section but not the s9 models.

\subsection{Comparison with \textsc{Vertex}}\label{sec:comp-with-vert}

The phases of collapse and early post-bounce evolution are simulated for all 1D and 2D models with the same input physics, thus we only need to compare these phases for models s20-ref-1D and s20VX-1D. The main features of the collapse are displayed in Fig.~\ref{fig:collapse}, which shows the electron-fraction, $Y_e$, lepton fraction, $Y_l$, and entropy per baryon, $s$, in the stellar center as functions of the time-evolving density, $\rho$, at the same location, as well as profiles of $Y_e$ and $s$ along the enclosed mass coordinate $m_{\mathrm{enc}}(r)=4\pi \int_0^r \rho \tilde{r}^2 \dd\tilde{r}$. The core deleptonization and heating, aided by neutrino-electron scattering \citep[e.g.][]{Bruenn1985, Thompson2003}, proceeds up to the trapping density of $\rho_c\sim 2\times 10^{12}\,$g\,cm$^{-3}$, above which $s$ and $Y_l$ remain essentially constant. The central values as well as the profiles agree well for both codes. The bounce\footnote{By definition, the bounce happens at the time when the shock entropy reaches 3\,$k_{\mathrm{B}}/$baryon.} commences $\approx 296$\,ms  (298\,ms) after initialization for the \textsc{Alcar} (\textsc{Vertex}) model at a mass coordinate of $m_{\mathrm{enc}}\approx0.57\,\Msol$ ($0.56\,\Msol$). In Fig.~\ref{fig:profiles} we show profiles of various quantities as functions of the enclosed mass (for times at and short after bounce) and radius (for later times). As can be seen in Fig.~\ref{fig:profiles}, the profiles of hydrodynamic quantities agree almost perfectly between \textsc{Alcar} and \textsc{Vertex} within the first tens of milliseconds after bounce. At this early stage the only noteworthy differences between both codes are observed in the neutrino-related quantities, e.g. in the luminosity,
\begin{equation}\label{eq:lumi}
   L \equiv r^2 \int_{\Omega,\eps} F_r \dd\eps\,\dd\Omega 
\end{equation}
and energy-averaged Eddington factor,
\begin{equation}\label{eq:eddfac}
  \langle\chi\rangle \equiv \frac{\int_{\Omega,\eps} \chi E \dd\eps\dd\Omega}{\int_{\Omega,\eps} E\dd\eps\dd\Omega} \, ,
\end{equation}
a few milliseconds after bounce right when the burst of electron neutrinos is released by the outward traveling shock wave. The aforementioned differences are related to what is seen in the small inset of the top left panel in Fig.~\ref{fig:timplot1d1}, namely that the $\nu_e$ burst in \textsc{Alcar} has $\sim 25\,$\% higher peak luminosity ($5.32\times10^{53}\,$erg\,s$^{-1}$ vs. $4.24\times10^{53}\,$erg\,s$^{-1}$) and contains $\sim 10\,$\% more energy ($2.79\times10^{51}\,$erg vs. $2.55\times10^{51}\,$erg as integrated between $t_{\mathrm{pb}}=0\,$ms and $10\,$ms) than the burst in \textsc{Vertex}. This difference goes hand in hand with a transiently lower deleptonization in the \textsc{Vertex} run (see Fig.~\ref{fig:profiles}, left column, third panel at $t_{\mathrm{pb}}=3\,$ms). The reason remains unknown; it might be related to the approximate two-moment closure for the transport employed by \textsc{Alcar}. The impact of the more energetic burst on the subsequent evolution is difficult to quantify, but it must be relatively small given that the agreement between both codes remains very good at later times. We also point out that \textsc{Vertex} exhibits a somewhat stronger diffusive broadening of the outward propagating $\nu_e$ burst compared to \textsc{Alcar}. This may be connected to a finer radial grid, smaller time step, and higher-order spatial reconstruction scheme used for integrating the transport equations in \textsc{Alcar}. This difference in the translatory behavior of the $\nu_e$ burst at large distances has no relevance for the model evolution.

Before comparing the post-bounce dynamics below the shock we take a look at the mass accretion rates measured at $r=500\,$km in Fig.~\ref{fig:macc}. The basically perfect match between \textsc{Alcar} and \textsc{Vertex} provides confidence that the numerical treatment of gravity and the tabulated EOS is handled consistently in both codes and, equally important, ensures that any differences of dynamics seen below the shock are not the result of different conditions above the shock.

In Fig.~\ref{fig:timplot1d1} we show important global quantities as functions of time for the four \textsc{Alcar} and \textsc{Vertex} models with and without $\nu-e^\pm$ scattering. The trajectory of the shock\footnote{Since the shock is not a perfect discontinuity but is distributed over several radial zones, we compute the shock radius as the arithmetic average of the radii bracketing the extended, numerical shock.}, $r_{\mathrm{s}}$, given in the top right panel of Fig.~\ref{fig:timplot1d1}, reveals that prompt shock expansion occurs for about 70-80\,ms, whereafter the shock stalls and recedes. Not surprisingly, in the present 1D models the shock retraction takes place smoothly and monotonically, except for a bump at $\tpb\sim 0.22-0.24\,$s that is related to the drop in the mass accretion rate (cf. Fig.~\ref{fig:macc}) due to the infalling Si/Si-O interface.

The neutrino luminosities, $L$, mean energies\footnote{We note that the mean energies computed as in Eq.~\eqref{eq:emean} are at $r=500\,$km essentially identical to the mean energies of the neutrino flux (obtained through replacing in Eq.~\eqref{eq:emean} $E$ by $F^r$ and $E/\eps$ by $F^r/\eps$) because $F^r/(c E)$ is very close to unity at this large radius.}
\begin{equation}\label{eq:emean}
  \langle\eps\rangle \equiv \frac{\int_{\Omega,\eps} E\dd\eps\,\dd\Omega}
  {\int_{\Omega,\eps} E\eps^{-1} \dd\eps\,\dd\Omega} \, 
\end{equation}
(both plotted in Fig.~\ref{fig:timplot1d1} at $r=500\,$km in the lab-frame as seen by an observer at infinity), the root-mean-squared (rms) energies,
\begin{equation}\label{eq:erms}
   \eps_{\mathrm{rms}} \equiv \sqrt{\frac{\int_{\Omega,\eps}\eps^2 E\dd\eps\,\dd\Omega}{\int_{\Omega,\eps} E\dd\eps\,\dd\Omega}} \,
\end{equation}
(plotted in Fig.~\ref{fig:timplot1d1} at the gain radius as seen by an observer locally comoving with the fluid), neutrino heating rates in the gain layer,
\begin{equation}
  Q_{\mathrm{heat}}\equiv \int_{\mathrm{gain}} Q_{\mathrm{E}} \dd V \, ,
\end{equation}
heating efficiencies,
\begin{equation}\label{eq:eta}
  \eta_{\mathrm{heat}}\equiv \frac{Q_\mathrm{heat}}{L_{\nu_e}+L_{\bar\nu_e}} \,
\end{equation}
(with luminosities defined as in Fig.~\ref{fig:timplot1d1}), as well as the characteristic timescales of neutrino heating and fluid advection through the gain layer (with mass $M_{\mathrm{g}}$ and sum of kinetic, internal, and gravitational energy of $E_{\mathrm{tot,g}}$),
\begin{subequations}
\begin{align}
  \tau_{\mathrm{heat}}\equiv |E_{\mathrm{tot,g}}|/Q_{\mathrm{heat}} \, , \\
  \tau_{\mathrm{adv}}\equiv M_{\mathrm{g}}/\dot{M}_{\mathrm{acc}} \, ,
\end{align}
\end{subequations}
respectively, agree for most of the time to within a few per cent between both codes\footnote{We adopt the same definition of the aforementioned quantities as \citet{Summa2016a}, except for possibly different locations and reference frames of measurement and the fact that the rms-energies here are based on the energy distribution, $E$, instead of the number distribution, $E/\eps$, such that the global heating rate can be approximately written as $Q_{\mathrm{heat}}\propto L\eps_{\mathrm{rms}}^2 M_{\mathrm{g}}/r_{\mathrm{g}}^2$, where $r_{\mathrm{g}}$ is the gain radius. Since only electron-type neutrinos contribute significantly to gain-layer heating, we omit plots and discussions of $\eps_{\mathrm{rms}}$ for $\nu_x$ neutrinos.}.

Within the small level of disagreement we notice the tendency of both \textsc{Alcar} models with and without neutrino-electron scattering to produce slightly higher neutrino luminosities, $L$, in both electron-neutrino species. This enhancement of energy-loss rates in \textsc{Alcar} may be the reason why we observe a slightly smaller radius, $r_{\mathrm{NS}}$, and higher temperature, $T_{\mathrm{NS}}$, of the PNS surface (defined by the location where $\rho=10^{11}\,$g\,cm$^{-3}$; cf. Fig.~\ref{fig:timplot1d1}), higher mean energies, $\langle\eps\rangle$, for all neutrino species, a less extended shock radius, $r_{\mathrm{s}}$, and gain radius, $r_{\mathrm{g}}$, as well as a higher mass-specific neutrino heating rate, $Q_{\mathrm{heat}}/M_{\mathrm{g}}$, for the \textsc{Alcar} models compared to the \textsc{Vertex} models. Remarkably, however, the ratio of the characteristic timescales, $\tau_{\mathrm{adv}}/\tau_{\mathrm{heat}}$ (shown in the bottom left panel in Fig.~\ref{fig:timplot1d1}), which is a more meaningful measure of the proximity to explosion than the aforementioned quantities \citep[e.g.][]{Janka2001a}, does not exhibit a clear trend in either direction. This means that the impact of a more compact configuration in \textsc{Alcar}, which in the first place is detrimental to an explosive runaway because the gain layer and shock sit deeper in the gravitational well, is approximately compensated by higher luminosities and therefore more powerful neutrino heating.

\subsection{Electron scattering}\label{sec:electron-scattering}

Figure~\ref{fig:timplot1d1} also shows the corresponding 1D results obtained when switching off neutrino-electron scattering at about $20\,$ms after bounce for both \textsc{Alcar} and \textsc{Vertex}. In the present 1D models the impact on many quantities is only on the percent level. Nevertheless, the less efficient energy deposition without neutrino-electron scattering (cf. $\eta_{\mathrm{heat}}$ in Fig.~\ref{fig:timplot1d1}) is sufficient to reduce the shock radius, $r_{\mathrm{s}}$, by a few km and to lead to a more sizable reduction of the advection timescale, $\tau_{\mathrm{adv}}$, and of the timescale ratio, $\tau_{\mathrm{adv}}/\tau_{\mathrm{heat}}$, by $10-20\,\%$. The most notable (but for the dynamics rather irrelevant) difference concerning the neutrino emission appears for the mean energies, $\langle\eps\rangle$, of the heavy-lepton neutrinos, which are several MeV higher without the efficient down-scattering process of neutrino-electron scattering (Fig.~\ref{fig:timplot1d1}, left column, second panel from top).

\subsection{Pair processes}\label{sec:pair-processes}

Figure~\ref{fig:timplot1d2} summarizes the evolution of the same quantities as in Fig.~\ref{fig:timplot1d1} but for the remaining one-dimensional \textsc{Alcar} models using the s20 progenitor.

We start by considering model s20-pp-1D that incorporates the simplified pair processes treatment, i.e. for electron-type neutrinos all pair processes are neglected and for  pair-annihilation of $\nu_x$ neutrinos the corresponding annihilation partners are assumed to be in isotropic LTE. The top left plot in Fig.~\ref{fig:timplot1d2} shows that the luminosities of heavy-lepton neutrinos are reduced compared to the reference case by $\sim 10\,\%$ during the first $\sim 150\,$ms of post-bounce evolution. The main reason for this reduction is most likely the approximate assumption that $\nu_x$ pair-annihilation targets are isotropically distributed in momentum space, whereas they actually become more and more forward peaked with increasing radius. This boosts the $\nu_x$ pair-annihilation rates while leaving the rates of the inverse (i.e. $\nu_x$ pair-production) reactions unchanged. We note that an impact of similar size has been found also by \citet{OConnor2015a} for 1D models with electron-positron annihilation but without bremsstrahlung. Keeping in mind that four times the individual $\nu_x$ luminosity enters the total energy loss rate of the PNS, this reduction of neutron-star cooling during $t_{\mathrm{pb}}\la 150\,$ms probably explains the observed increase of the neutron-star radius by $\sim 1-2\,$km and of the shock radius by $\sim 3-10\,$km compared to the reference model.

The luminosities and spectral properties of emitted electron-type neutrinos remain fairly unaffected by the pair-process simplification. This helps understanding the similarly weak sensitivity of the specific heating rate, which can approximately be written as
\begin{equation}
  \frac{Q_{\mathrm{heat}}}{M_{\mathrm{g}}} \propto \frac{L\eps_{\mathrm{rms}}^2}{r_{\mathrm{g}}^2}
\end{equation}
(where $L\eps_{\mathrm{rms}}^2\equiv L_{\nu_e} \eps_{\mathrm{rms,\nu_e}}^2 +L_{\bar\nu_e} \eps_{\mathrm{rms,\bar\nu_e}}^2$), and the heating timescale, which roughly scales like
\begin{equation}
  \tau_{\mathrm{heat}}\propto  \frac{|E_{\mathrm{tot,g}}|}{M_{\mathrm{g}}}
  \frac{r_{\mathrm{g}}^2}{L\eps_{\mathrm{rms}}^2} \propto \frac{G\,M_{\mathrm{NS}} r_{\mathrm{g}}}{L\eps_{\mathrm{rms}}^2}
\end{equation}
(where we assumed the specific total energy of the gain region to be approximately proportional to the gravitational energy at the gain radius). The heating rate, heating efficiency and the advection timescale, on the other hand, show stronger deviations from the reference model, mainly because they are directly proportional to the mass in the gain layer, which itself is enhanced by $\sim 10-20\,\%$ compared to the reference model. As a result, $\tau_{\mathrm{adv}}/\tau_{\mathrm{heat}}$, which is most relevant for the explosion behavior, is enhanced by a comparable amount during almost the entire evolution.

We note that this result is not in tension with previous studies (and with the results for models s20-str-1D and s20-mb-1D discussed below) that report more favorable runaway conditions for cases in which the PNS contraction proceeds faster \citep[e.g.][]{Marek2009a,OConnor2018a, Melson2015b, Bollig2017a}. This is because in those studies the faster PNS contraction is the result of changing physics ingredients different from the ones varied here, e.g, using a softer nuclear equation of state, general relativity instead of Newtonian gravity, or reduced nucleon-scattering opacities for all neutrinos. In those cases the accelerated PNS contraction came along with more favorable neutrino emission properties (i.e. higher values of $L\eps_{\mathrm{rms}}^2$) that overcompensated for the stronger gravitational binding. In the present case, on the other hand, the deceleration of PNS contraction due to artificially amplified annihilation of $\nu_x$ pairs is not accompanied by a large enough reduction of $L\eps_{\mathrm{rms}}^2$ to cause more pessimistic runaway conditions.

\subsection{Strangeness and many-body corrections}\label{sec:strang-many-body}

The two 1D models, s20-str-1D and s20-mb-1D, both incorporate corrections that effectively reduce the opacities for neutrino-nucleon scattering, in particular in the crucial neutrino-decoupling region, by a few percent. As a result, we observe an enhancement of the neutrino luminosities for all three species by a similar amount, which is well in agreement with previous studies that incorporated these corrections \citep[e.g.][]{Melson2015b, OConnor2017a}. In the present one-dimensional case, this leads to smaller shock- and PNS-radii throughout, but slightly higher mean- and rms-energies of neutrinos and therefore, in combination with the increased luminosities, to slightly stronger neutrino heating, shorter $\tau_{\mathrm{heat}}$ and higher $\tau_{\mathrm{adv}}/\tau_{\mathrm{heat}}$, thus slightly more favorable conditions for shock revival.

\subsection{Velocity-dependent and gravitational redshift terms}\label{sec:veloc-depend-grav}

\begin{figure*}
\centering
 \includegraphics[width=0.33\textwidth]{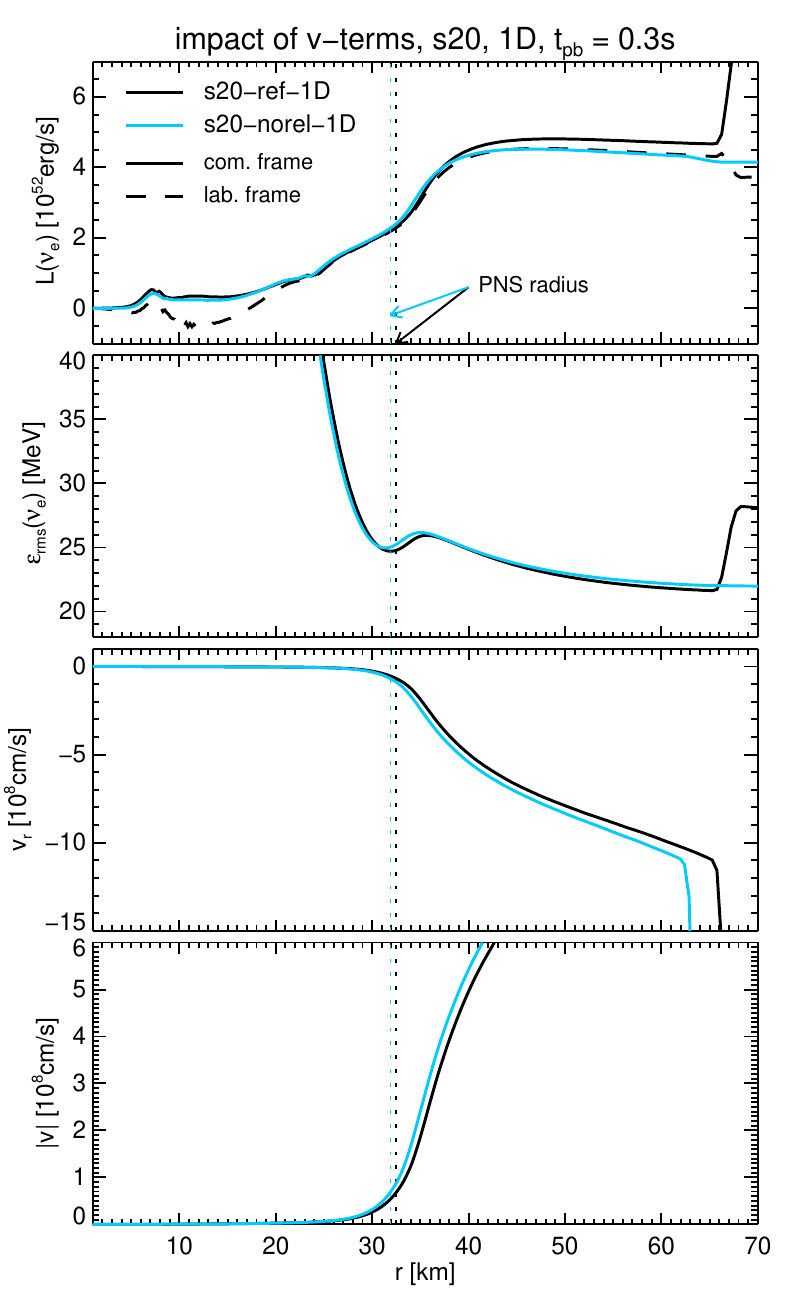}
 \includegraphics[width=0.33\textwidth]{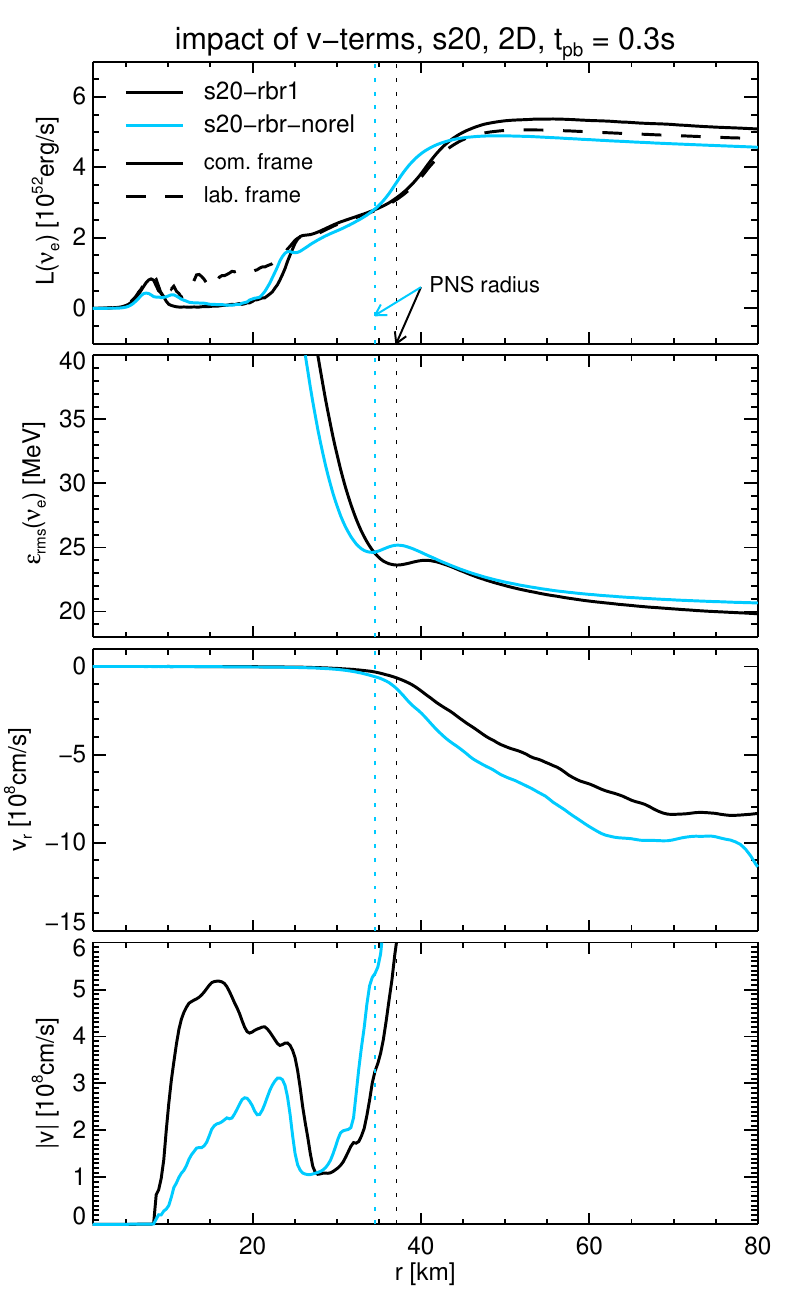}
 \includegraphics[width=0.33\textwidth]{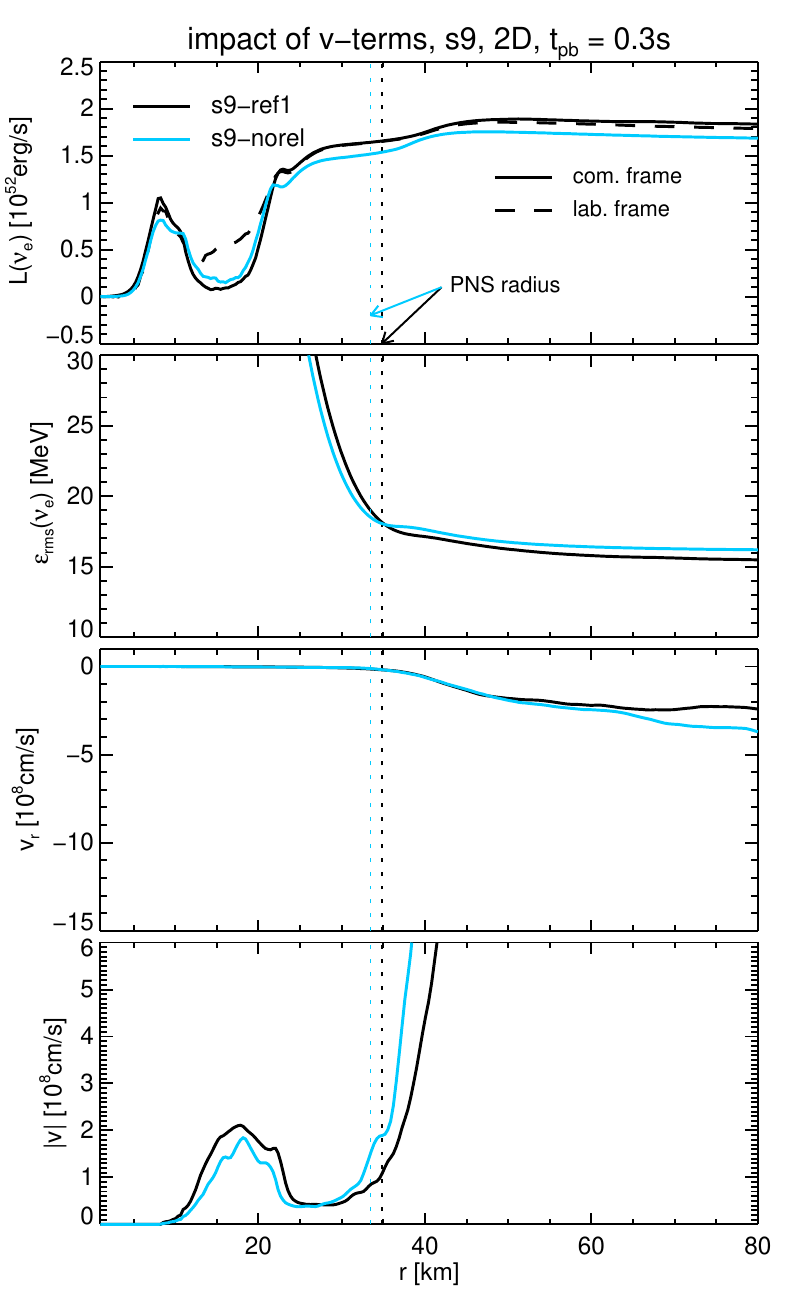}
 \caption{Comparison between simulations that include (black lines) and neglect (light blue lines) velocity-dependent and gravitational redshift terms in the transport for spherically symmetric s20 models (left panels), axisymmetric s20 models (middle panels), and axisymmetric s9 models (right panels). Shown are from top to bottom radial profiles of luminosities and rms-energies of electron neutrinos $\nu_e$, as well as angle averages of the radial velocities and absolute velocities. The plotted data have been averaged over 20\,ms around the displayed times. Neutrino-related quantities displayed with solid lines have been computed using the evolved neutrino moments, while dashed lines in the top panels show (for models including velocity-dependent terms) the luminosities transformed into the local (i.e. as measured by an observer at the corresponding radius) lab-frame. Vertical dotted lines indicate the PNS radius, $r_{\mathrm{NS}}$, where by definition $\rho=10^{11}\,$g\,cm$^{-3}$.}
\label{fig:vtermspns}
\end{figure*}	

For model s20-norel-1D, in which frame-dependent effects such as advection and Doppler-shift as well as gravitational redshift are ignored after $t_{\mathrm{pb}}\approx20\,$ms, one immediately recognizes the significantly smaller heating rate as well as specific heating rate, reduced heating efficiency, longer heating timescale, and reduced timescale ratio compared to the reference case. At first glance, this seems counterintuitive in view of the fact that according to Fig.~\ref{fig:timplot1d2} the (lab-frame) luminosities agree well and the (lab-frame) mean energies are even higher compared to the reference model. However, both quantities are not sufficiently representative of the heating conditions in the gain layer, as is revealed by Fig.~\ref{fig:vtermspns} that depicts radial profiles of the luminosities and rms-energies measured in the comoving frame\footnote{For model s20-norel-1D the distinction between the comoving frame and lab frame is meaningless and the evolved neutrino moments are used for plots of both comoving- and lab-frame quantities.} at a representative post-bounce time for both 1D and 2D models: For model s20-norel-1D the neutrino fluxes as seen in the fluid frame are almost $10\,\%$ smaller in the gain region than for the reference model. The good agreement between the luminosities of s20-norel-1D with the local lab-frame luminosities of the reference model (dashed lines in Fig.~\ref{fig:vtermspns}) suggests that this difference stems from Doppler boosting in the infalling material, which is ignored in model s20-norel-1D. One might still wonder why the rms-energies, $\eps_{\mathrm{rms}}$ (cp. Figs.~\ref{fig:timplot1d2} and~\ref{fig:vtermspns}) exhibit hardly any reduction in model s20-norel-1D compared to the reference model. The most likely explanation is that the frame correction is in relative terms smaller for the rms-energies (transforming as $\eps^{\mathrm{lab}} \approx \eps^{\mathrm{comoving}}(1+v/c)$ for forward peaked radiation) than for the luminosities (transforming as $L^{\mathrm{lab}} \approx L^{\mathrm{comoving}}(1+2v/c)$ for forward peaked radiation), such that gravitational redshift may approximately compensate for the effect of Doppler blueshift for the rms-energies.

\section{Results: 2D models}\label{sec:2d-axisymm-models}

After having obtained an idea of the level of agreement between the two codes \textsc{Alcar} and \textsc{Vertex} and the impact of our modeling variations for the spherically symmetric case, we now investigate how these dependencies translate to the 2D axisymmetric case. Additionally, we will examine the impact of using the RbR+ approximation for the \textsc{Alcar} code, comment on the level of stochasticity and numerical convergence, and contrast our results for the s20 models with others found in the literature.

\subsection{Basic features}\label{sec:main-evol-feat}

Before comparing the models in detail, we first summarize some features common to most models. Figure~\ref{fig:contours} depicts snapshots of the isotropic-equivalent lab-frame luminosity, $4\pi r^2 \int_\eps F^{\mathrm{lab}}\dd\eps$, and the specific entropy taken at three post-bounce times for two \textsc{Alcar} models and one \textsc{Vertex} model, and Figs.~\ref{fig:timplot2d1},~\ref{fig:timplot2d2}, and~\ref{fig:timplot2d_s9} provide a summary of global properties for most investigated models. By switching to 2D, we allow the fluid to develop fluid instabilities such as PNS convection, SASI, and post-shock convection driven by neutrino heating.

Proto-neutron star convection is the consequence of unstable gradients of the lepton number and specific entropy developing below the PNS surface as a result of neutrino emission. Leptons trapped in the dense core of the PNS are shuffled into the overlying layers, where they provide additional pressure support and thereby slow down the PNS contraction. Correspondingly, we observe larger PNS radii, $r_{\mathrm{NS}}$, and lower temperatures, $T_{\mathrm{NS}}$, as well as reduced energies of emitted neutrinos, $\langle\eps\rangle$ and $\eps_{\mathrm{rms}}$, in 2D compared to 1D at equal times \citep[][]{Buras2006,Dessart2006}. The luminosities, $L$, of electron-type neutrinos remain almost unchanged, while those of the heavy-lepton neutrinos are enhanced by several tens of percent. All our models (although with some differences for those models neglecting velocity terms in the transport, see Sec.~\ref{sec:velocity-terms}), reproduce these effects connected to PNS convection in good agreement with previous studies \citep[e.g.][]{Buras2006, Dessart2006, Muller2012b, Bruenn2016a, Radice2017a}.

While PNS convection already starts at about $20-40$\,ms after bounce, the gain layer remains fairly spherically symmetric for a much longer time, namely until about $t_{\mathrm{pb}}\sim 0.15$\,s ($\sim 0.1\,$s) for the s20 (s9) progenitor models, as can be recognized in Figs.~\ref{fig:timplot2d1} and~\ref{fig:timplot2d2} (\ref{fig:timplot2d_s9}) for the s20 (s9) models by the correspondingly late rise of the lateral (i.e. carried by non-radial motions) kinetic energy integrated over the gain layer, $E_{\mathrm{kin,g}}^{\mathrm{lat}}$. This means that until these times the conditions for efficient growth are met neither for SASI nor for post-shock convection. Since the hydrodynamic instabilities are triggered by random perturbations, the exact time of significant departure from spherical symmetry slightly varies between the models. It may be worth pointing out that in tests with \textsc{Alcar} using the s20 model we experienced quite some sensitivity of this transition time to the numerical treatment: Using a low (i.e. linear) order for spatial reconstruction of the hydro variables or employing the more diffusive HLL solver led to a significantly earlier onset of non-radial flow activity.

\subsubsection{s20 progenitor models}\label{sec:s20-prog-models}

Since SASI growth is favored by short advection timescales \citep{Blondin2006, Foglizzo2007a, Scheck2008}, $\tau_{\mathrm{adv}}$, efficient growth sets in once the post-shock configuration has become sufficiently compact. Around $150-200$\,ms after bounce the exponential growth of SASI modes with periods comparable to $\tau_{\mathrm{adv}}$ can be seen in the dipole component of the shock surface (cf. bottom left panel of Figs.~\ref{fig:rbrcomp_s20}), which is obtained from a decomposition of the latter into Legendre polynomials \citep[see, e.g.,][for the exact definition of the multipole coefficients]{Summa2016a}. The snapshots in the top row of Fig.~\ref{fig:contours} are taken at $t_{\mathrm{pb}}\sim 180\,$ms right around the time when the transition from the well-ordered, linear phase to the turbulent, non-linear phase of the SASI takes place. After this transition the evolution of the post-shock layer is characterized by parasitic instabilities (Rayleigh-Taylor and Kelvin-Helmholtz) whose turbulent mass motions tap energy from the SASI. Post-shock convection tends to be suppressed for non-exploding
s20 runs because of the retreating shock radius and correspondingly short advection timescales, which act against the development of a sufficiently large negative entropy gradient and correspondingly short growth timescales for convection. This statement is backed by the circumstance that the $\chi_{\mathrm{conv}}$-parameter characterizing the growth conditions of post-shock convection (see, e.g., \citealp{Foglizzo2006a, Summa2016a} for the definition of this quantity and details regarding its interpretation and computation) remains below the critical value of $\approx 3$ (bottom right panel of Figs.~\ref{fig:timplot2d1} and~\ref{fig:timplot2d2}). Nevertheless, since the critical condition for convection, $\chi_{\mathrm{conv}}>3$, holds strictly only for the linear regime, the existence of secondary convective instability associated with highly non-linear SASI activity is not excluded.

\subsubsection{s9 progenitor models}

\begin{figure*}
\centering
 \includegraphics[width=\textwidth]{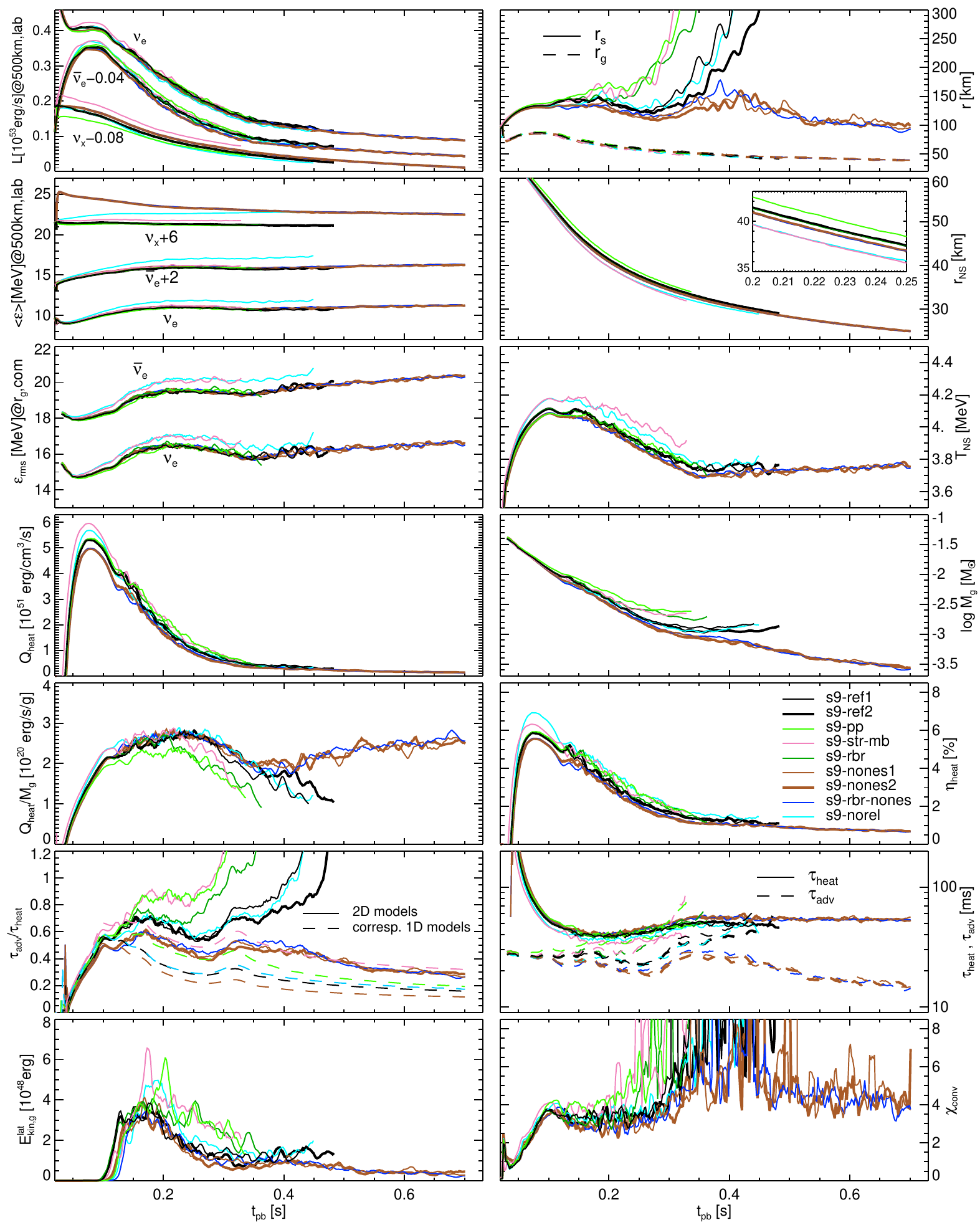}
 \caption{Same as Fig.~\ref{fig:timplot1d1} but for the axisymmetric s9 models whose names are displayed in the panel for $\eta_{\mathrm{heat}}$, and additionally providing the lateral kinetic energies in the gain region, $E^{\mathrm{lat}}_{\mathrm{kin,g}}$, as well as the convection parameter, $\chi_{\mathrm{conv}}$, in the bottom row. The suite of models compares variations of the input physics to the reference cases of the $9\,\Msol$ \textsc{Alcar} simulations, s9-ref1 and s9-ref2. The curves are smoothed using running averages of 10\,ms.}
\label{fig:timplot2d_s9}
\end{figure*}

For the models using the s9 progenitor, which are characterized by considerably smaller mass accretion rates with respect to the s20 models (see Fig.~\ref{fig:macc}) and therefore a less compact shock surface, the situation is reversed in that the dominant fluid instability is not SASI but post-shock convection. Correspondingly, we observe a transition of the $\chi_{\mathrm{conv}}$-parameter above the critical value of $\sim 3$ at about $t_{\mathrm{pb}}\sim 0.1$\,s (cf. Fig.~\ref{fig:timplot2d_s9}), whereafter $\chi_{\mathrm{conv}}$ remains $\ga 3$ during the entire simulation. This strong progenitor dependence of instability regimes, with massive (low-mass) progenitors favoring SASI (convection) has been recognized before, e.g. in \citet{Muller2012a, Fernandez2014a}.

\subsection{Comparison with \textsc{Vertex}}\label{sec:comp-with-vert-1}

\begin{figure*}
  \centering 
  \includegraphics[width=\textwidth]{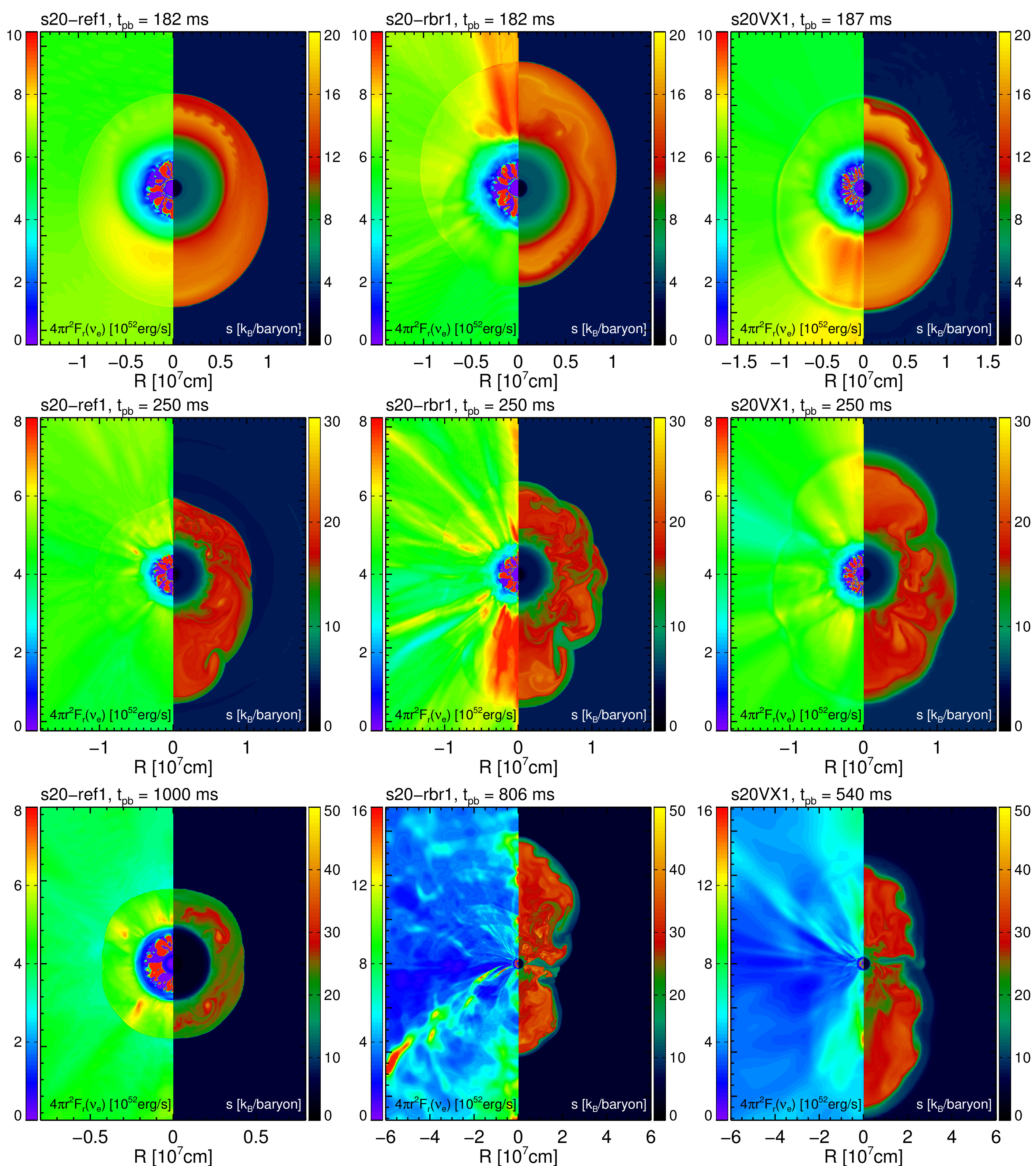}
  \caption{2D maps of the isotropic-equivalent luminosities, $4\pi r^2\int_\eps F_r\dd\eps$, measured in the local lab-frame (left side of each panel) and specific entropy, $s$ (right side of each panel), for the \textsc{Alcar} reference model using the s20 progenitor (s20-ref1; left column), the RbR+ counterpart of the latter (s20-rbr1; middle column), and the corresponding \textsc{Vertex} model (s20VX1; right column) at an early time during the transition from the linear to the non-linear phase of the SASI (top row), at an intermediate time right after the infall of the Si/Si-O interface (middle row), and at a late time showing the failed or successful onset of shock runaway (bottom row).}
  \label{fig:contours}
\end{figure*}

\begin{figure*}
\centering
 \includegraphics[width=\textwidth]{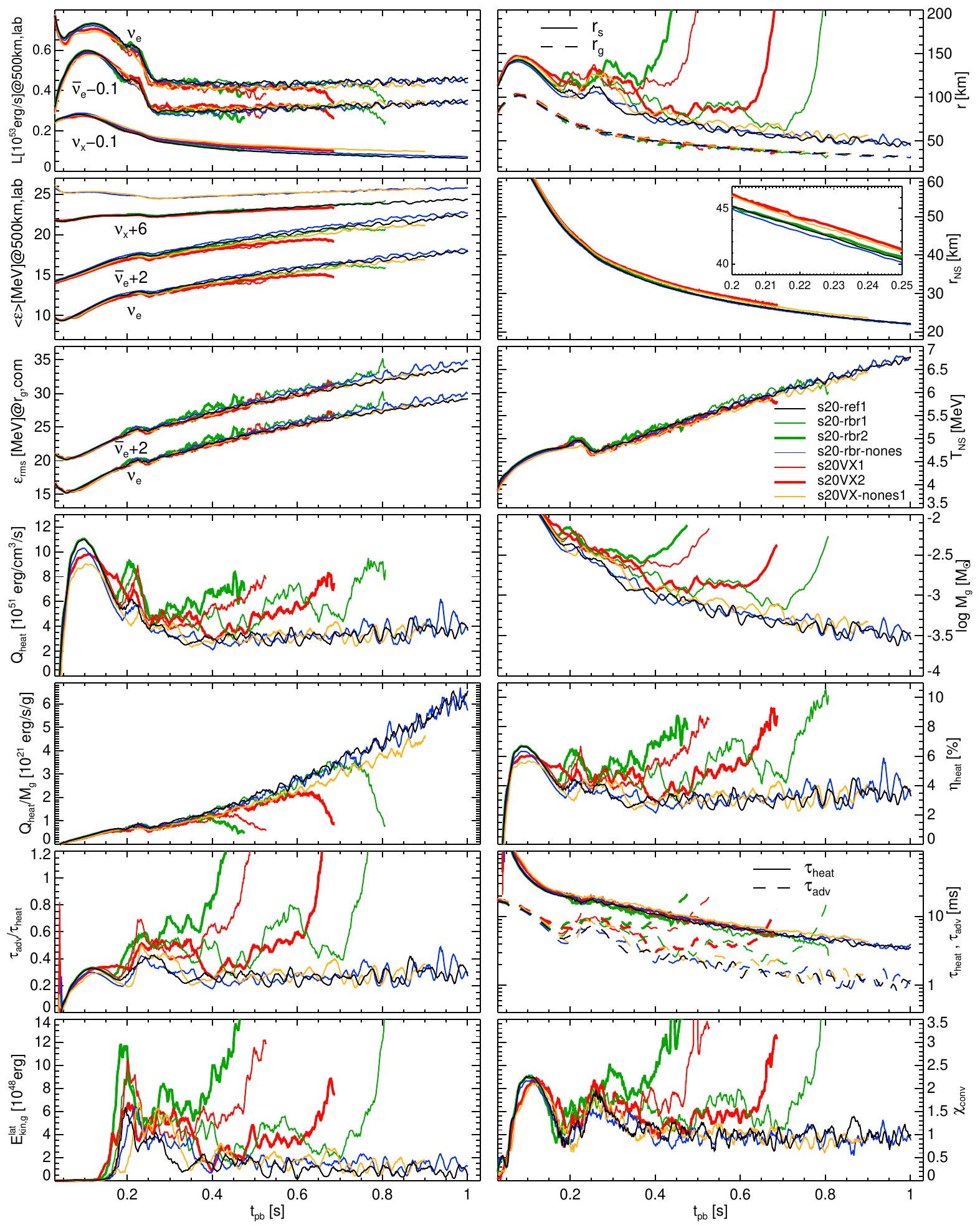}
 \caption{Same as Fig.~\ref{fig:timplot1d1} but for the axisymmetric s20 models whose names are displayed in the panel for $T_{\mathrm{NS}}$, and additionally providing the lateral kinetic energies in the gain region, $E^{\mathrm{lat}}_{\mathrm{kin,g}}$, as well as the convection parameter, $\chi_{\mathrm{conv}}$, in the bottom row. The compilation of models compares the RbR+ cases with \textsc{Alcar} and \textsc{Vertex} to the reference 2D model s20-ref1. Lines with enhanced thickness but same color denote models with same corresponding physics ingredients but a different initial perturbation pattern. The curves are smoothed using running averages of 10\,ms.}
\label{fig:timplot2d1}
\end{figure*}	

\begin{figure*}
  \centering 
  \includegraphics[width=0.49\textwidth]{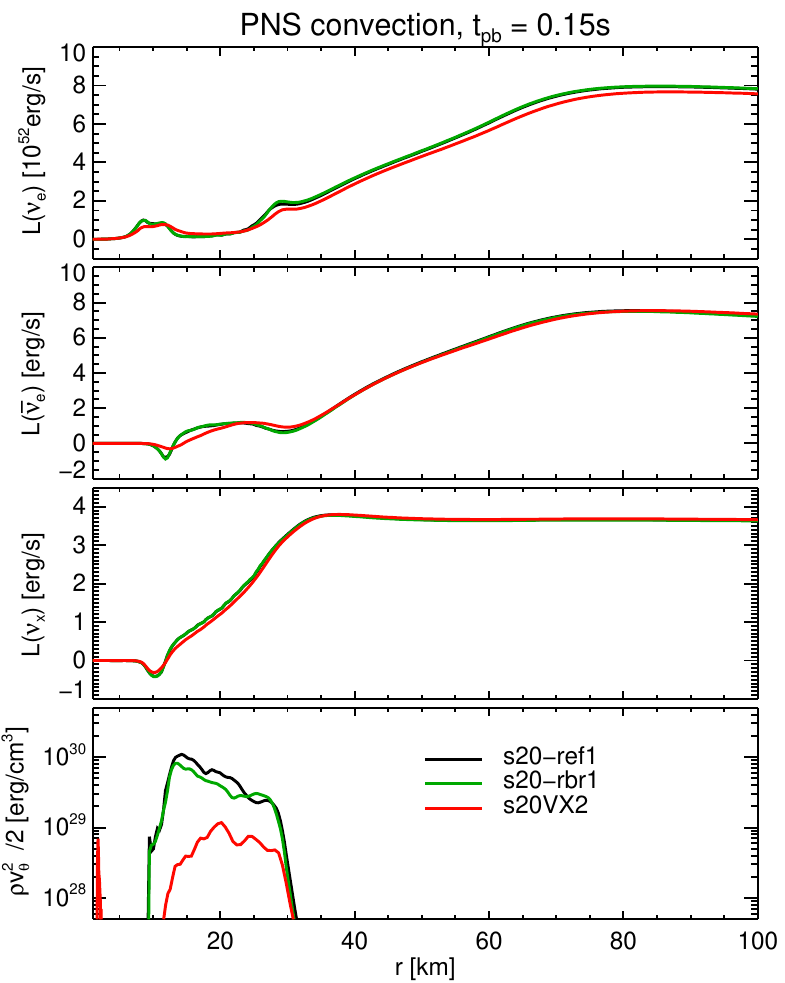}
  \includegraphics[width=0.49\textwidth]{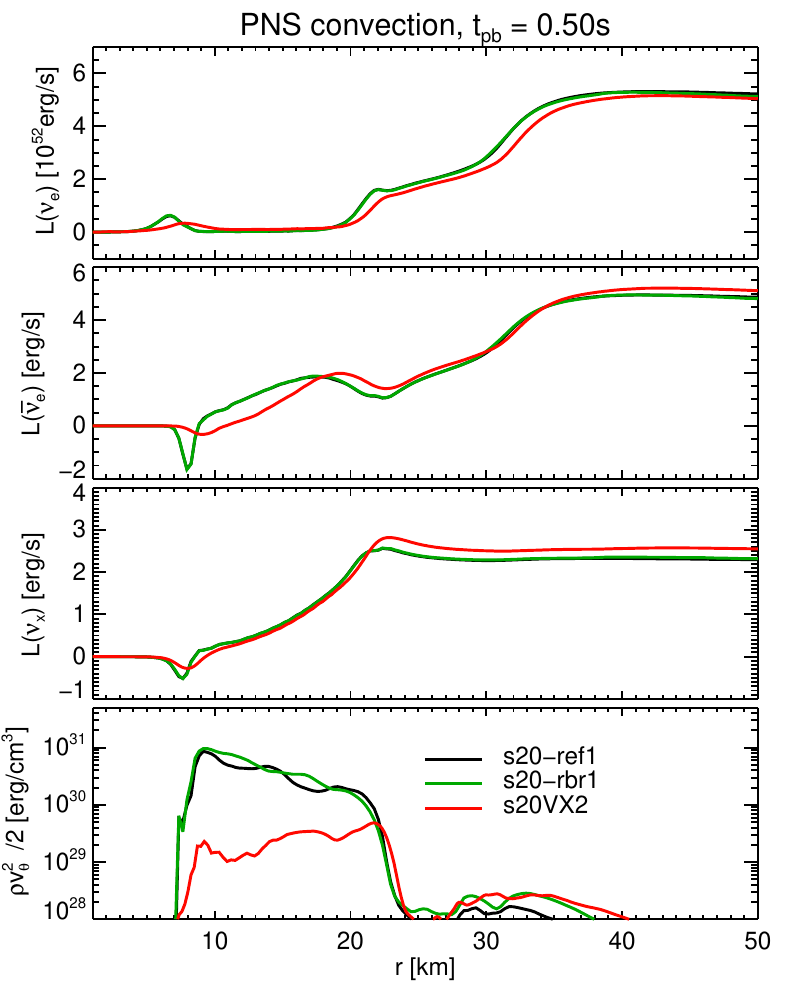}
  \caption{Comoving-frame luminosities of the three evolved neutrino species ($\nu_e$, $\bar\nu_e$, and $\nu_x$, from top panels downwards) as well as angle averaged lateral kinetic energies, $\rho v_\theta^2/2$ (bottom panel), for the reference \textsc{Alcar} model, s20-ref1, its pendant using the RbR+ approximation, s20-rbr1, and the corresponding \textsc{Vertex} model, s20VX2, for two representative times (left and right columns). The plotted data have been averaged over 20\,ms around the displayed times. Proto-neutron star convection appears to be insensitive to the use of RbR+, but it is more energetic in \textsc{Alcar} than in \textsc{Vertex}. The impact of this discrepancy on the neutrino fluxes leaving the PNS surface remains small but is increasing in time as the PNS evolves and evolutionary differences accumulate.}
  \label{fig:pnscomp}
\end{figure*}	

We start by comparing the two \textsc{Alcar} models that incorporate the RbR+ approximation with (s20-rbr) and without (s20-rbr-nones) neutrino-electron scattering to the corresponding \textsc{Vertex} runs (s20VX and s20VX-nones, respectively).

First of all, as visible in Fig.~\ref{fig:timplot2d1} and Table~\ref{table_models}, the models without neutrino-electron scattering do not explode (at least not until the end of each simulation), while the ones including neutrino-electron scattering do explode, but rather late (i.e. several hundred milliseconds after the infall of the Si/Si-O interface at $\sim 230\,$ms). The agreement of both codes in clearly showing the impact of a relatively small $\mathcal{O}(5\,\%)$ variation of neutrino-interaction rates (in the present case due to neutrino-electron scattering) is thus already encouraging. Moreover, for both codes the exploding models are characterized by a significant scatter in explosion times (cf. Table~\ref{table_models}): The three \textsc{Alcar} models with the same input physics but different random initial perturbation patterns cover a large range of explosion times of $t_{\mathrm{exp}}= 0.48-0.92\,$s, while the two corresponding \textsc{Vertex} runs explode also within this time interval, namely at $t_{\mathrm{exp}}= 0.53\,$s and $0.7\,$s.

The overview of important global properties as functions of time in Fig~\ref{fig:timplot2d1} reveals that the differences between both codes concerning the neutrino luminosities and energies, as well as the PNS surface temperature and radius remain on the few-percent level, as found in 1D. This suggests that PNS convection is operating in both codes consistently regarding its impact on PNS contraction and neutrino emission. Nevertheless, a closer look unfolds that the $\bar\nu_e$ and $\nu_x$ luminosities now show a small enhancement in \textsc{Vertex} relative to \textsc{Alcar} for $t_{\mathrm{pb}}\ga 0.15\,$s that was not seen in 1D (see Fig.~\ref{fig:timplot1d1}) and may therefore be connected to some discrepancy in the PNS convection. Indeed, an inspection of the radial profiles of the lateral kinetic energy densities and the neutrino luminosities in Fig.~\ref{fig:pnscomp} reveals that the PNS convection in \textsc{Alcar} proceeds somewhat differently with higher kinetic energies. Unfortunately, despite a dedicated analysis we were unable to track down the exact reason for this discrepancy and its detailed consequences. We can, however, already clearly assess that this enhancement in convective energy is not an artifact of the RbR+ approximation, because model s20-ref1, which does not employ RbR+, shows a similar behavior (cf. Fig.~\ref{fig:pnscomp} and a more detailed discussion of the RbR+ approximation in Sec.~\ref{sec:ray-ray-plus}). It may be that the stronger PNS convection in \textsc{Alcar} is related to a subtle difference in the PNS structure, which is systematically more compact ($\sim 1\,$km difference in $r_{\mathrm{NS}}$) than in \textsc{Vertex} models in 1D (Fig.~\ref{fig:timplot2d1}) as well as 2D. This might allow PNS convection to be enhanced in \textsc{Alcar} by tapping the higher gravitational binding energy of the more compact PNS.

We do not deem the disagreement in the PNS convection zone to be overly significant concerning the explosion dynamics, because the shock radius, the neutrino heating rates, characteristic timescales, as well as most remaining properties are in good agreement apart from stochastic fluctuations.

Assessing in detail the heating conditions in the gain layer and the shock evolution, and ultimately finding the reason why a simulation at a certain time exhibits shock expansion or not, is difficult given the complicated temporal behavior of heating-related diagnostic quantities such as  $Q_{\mathrm{heat}}$, $\eta_{\mathrm{heat}}$, $E^{\mathrm{lat}}_{\mathrm{kin,g}}$, and $\tau_{\mathrm{adv}}/\tau_{\mathrm{heat}}$. At later times, $t_{\mathrm{pb}}\ga 0.3$\,s, we observe, consistently for both codes, that the exploding models in comparison to the non-exploding models exhibit temporal variations on longer timescales and with larger amplitudes. For instance, $\tau_{\mathrm{adv}}/\tau_{\mathrm{heat}}$, occasionally travels up and down between $\sim 0.3-0.6$ on timescales of hundreds of milliseconds for the exploding models, while for the non-exploding models it remains rather close to $\sim 0.3$ with relatively short and low-amplitude oscillations. The larger amplitudes and longer timescales of temporal variations suggest that the exploding models hover in a state very close to criticality before they ultimately explode. Retaining in such a state means that small perturbations may have a large dynamic impact. A natural suggestion from this is that the temporal pattern observed for the heating-related properties is shaped more by stochasticity than systematics. This conjecture is supported when comparing models that were initialized with a different pattern of density perturbations (cp. thick and thin lines of same color in Fig.~\ref{fig:timplot2d1}): $\tau_{\mathrm{adv}}/\tau_{\mathrm{heat}}$, for instance,
varies over a comparable range of values but with a considerably different temporal behavior for different initial perturbations. We infer that the time of shock runaway must be similarly affected by stochasticity, which is confirmed by the large scatter of runaway times seen for both codes. Therefore, even though the explosion times of individual simulations may differ by several hundred milliseconds, we conclude that the overall agreement between both codes is good and no worrisome differences concerning the explosion behavior are found.

Before moving on to the next topic, we comment on a rather peculiar difference between \textsc{Alcar} and \textsc{Vertex}, namely that in \textsc{Alcar} the evolved neutrino energies and fluxes appear to have rather fine spatio-temporal structures while the neutrino quantities are smoother in \textsc{Vertex} (see, e.g., the color maps of the isotropic-equivalent luminosities in Fig.~\ref{fig:contours}). We suspect two possible reasons: First, the approximate, non-linear closure in \textsc{Alcar}, which allows self-interaction and non-linear effects (such as shocks) of the radiation moments even in optically thin regions \citep[e.g.][]{Pons2000}, and second, the use of an explicit time integrator in \textsc{Alcar} as opposed to an implicit one in \textsc{Vertex} that tends to broaden sharp features. At this point it is thus not clear up to which level the observed small-scale features in \textsc{Alcar} are physical or not; however, the dynamical consequences seem to be marginal given the otherwise good agreement.

\begin{figure*}
\centering
\includegraphics[width=\textwidth]{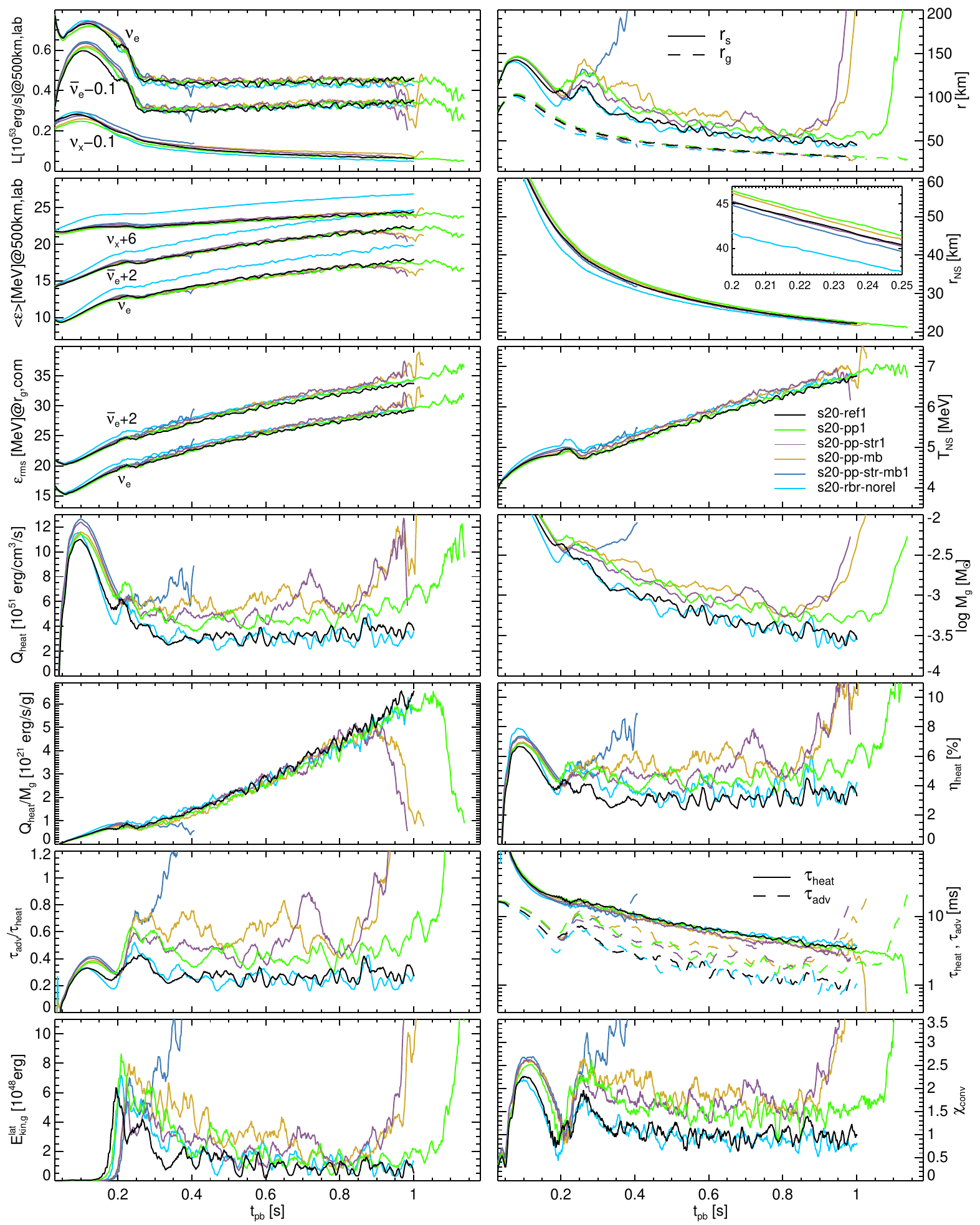}
\caption{Same as Fig.~\ref{fig:timplot1d1} but for the axisymmetric s20 models whose names are displayed in the panel for $T_{\mathrm{NS}}$, and additionally providing the lateral kinetic energies in the gain region, $E^{\mathrm{lat}}_{\mathrm{kin,g}}$, as well as the convection parameter, $\chi_{\mathrm{conv}}$, in the bottom row. The compilation of models compares different variations of the physics input of the \textsc{Alcar} simulations with the reference model s20-ref1. The curves are smoothed using running averages of 10\,ms.}
\label{fig:timplot2d2}
\end{figure*}	

\subsection{Electron scattering, pair processes,  strangeness  and many-body corrections}\label{sec:electr-scatt-pair}

As already pointed out in the previous section the relatively small effects associated with neutrino-electron scattering, which has often been neglected in a number of previous studies assuming that it would only be relevant during collapse, turn out to be a crucial ingredient for initiating shock runaway for the s20 model \citep[in qualitative agreement with][]{Burrows2018a}. For the s9 progenitor models (see Table~\ref{table_models} and Fig.~\ref{fig:timplot2d_s9} for an overview of global properties as functions of time) we see the same strong sensitivity, i.e. models exploding around $t_{\mathrm{pb}}\sim 0.3-0.5\,$s (s9-ref, s9-rbr) are turned into non-exploding models (s9-nones, s9-rbr-nones) when ignoring neutrino-electron scattering. These results demonstrate once more how sensitive the onset of shock runaway can be to variations in the neutrino interaction rates.

In view of the substantial impact of neutrino-electron scattering it seems not too astonishing that we see a similarly strong impact of other variations of the neutrino rates considered in this study, both for the s20 (see Fig.~\ref{fig:timplot2d2}) and s9 (see Fig.~\ref{fig:timplot2d_s9}) models: The simplified treatment of pair processes (used in all models labeled by ``pp'' in their names), as well as the scattering-opacity reducing strangeness and many-body corrections (including ``str'' and ``mb'' in their names, respectively) are all conducive to explosions, i.e. each modification comes along with a visible boost of the timescale ratio, $\tau_{\mathrm{adv}}/\tau_{\mathrm{heat}}$, as well as (in the case of a successful explosion) a shift towards earlier times of shock expansion. This shift is naturally larger in the s20 models, because the reference s20 model (s20-ref) lacks an explosion while the reference s9 model (s9-ref) already explodes around $t_{\mathrm{pb}}\sim 0.4$\,s. Combining all three of these variations turns a robustly (in the sense of being independent of stochasticity) non-exploding model (s20-ref) into a robustly exploding model (s20-pp-str-mb), with explosion times between 0.38-0.50\,s, cf. Table~\ref{table_models}.

Crudely judging from the explosion times (cf. Table~\ref{table_models}), the impact of the
pair-process simplification is roughly comparable to that of the combination of the two physically motivated opacity corrections (``str'' and ``mb''; cp. models s9-ref, s9-str-mb, and s9-pp) and slightly stronger than that of neutrino-electron scattering (cp. models s20-rbr-pp, s20-rbr, and s20-rbr-pp-nones).

Finally, we point out an interesting feature: The impact of each microphysics variation on the explosion behavior is qualitatively correctly predicted in 1D by the timescale ratio, $\tau_{\mathrm{adv}}/\tau_{\mathrm{heat}}$, in the sense that all modifications, which lead to larger values of $\tau_{\mathrm{adv}}/\tau_{\mathrm{heat}}$ in 1D, trigger an earlier onset of explosion, or an explosion at all, in 2D (see for the s20 models Figs.~\ref{fig:timplot1d2} and Sec.~\ref{sec:results:-1d-models}, as well as for the s9 models the dotted lines in the panel for $\tau_{\mathrm{adv}}/\tau_{\mathrm{heat}}$ in Fig.~\ref{fig:timplot2d_s9} showing the behavior of the corresponding 1D models). In contrast, the shock trajectory of 1D models is not a good indicator of more/less favorable runaway conditions: Both strangeness and many-body corrections reduce the shock radius in 1D, while the pair-process simplification increases the latter (cf. Fig.~\ref{fig:timplot1d2}), but all three of these physics variations help initiating a shock runaway in 2D.

\subsection{Velocity-dependent and gravitational redshift terms}\label{sec:velocity-terms}

Concerning the prospects of shock runaway, the purely negative impact of neglecting velocity-dependent as well as gravitational redshift terms in the transport seen in 1D seems to be partially alleviated in 2D, albeit in a model-dependent fashion: An early exploding
s20 model (s20-pp-str-mb) explodes even earlier with these modifications (s20-pp-str-mb-norel), a late exploding s20 model (s20-rbr) then lacks an explosion (s20-rbr-norel), and the reference s9 model (s9-ref) seems almost unaffected (s9-norel).

These results are most likely connected to the reduced convective activity in the PNS, which can be inferred from the smaller values of absolute velocities, $|v|$, in the PNS convection zone displayed for a representative time in Fig.~\ref{fig:vtermspns}. The consequence is a PNS that contracts faster, is hotter, and therefore emits neutrinos with higher rms-energies (see $r_{\mathrm{NS}}$, $T_{\mathrm{NS}}$, and $\eps_{\mathrm{rms}}$ in Figs.~\ref{fig:timplot2d1},~\ref{fig:timplot2d2}, and~\ref{fig:timplot2d_s9}). For a less massive PNS these explosion facilitating consequences seem to be able to (over-)compensate for the missing Doppler blueshift in the infalling material, because the infall velocities are lower compared to a more massive PNS (see profiles of the radial velocity, $v_r$, in Fig.~\ref{fig:vtermspns}). This dependence on the PNS mass at the time of close proximity to criticality might explain why the runaway conditions become more (less) optimistic for the originally early (late) exploding model when switching off the considered terms. In the models with the less massive s9 progenitor, all velocities, including those in the PNS convection zone, are lower to begin with, which might explain the almost vanishing net impact on the shock trajectory in this case.

The actual reason for the less efficient PNS convection when switching off the considered terms is not determined easily. We suspect the following: When neglecting velocity-dependent terms in the moment equations, neutrinos (i.e. their energy and lepton number) are not advected with the flow as they should, being deep inside the PNS. Instead, neutrinos carried in a bubble that is about to rise in response to an unstable stratification effectively leak out of the bubble and are left behind. Since near the bottom of the PNS convection zone the neutrino lepton number can be a sizable fraction of $Y_e$, these neutrino losses effectively reduce the lepton number of the bubble and therefore the tendency of the latter to rise further.

\begin{figure*}
  \centering
 \includegraphics[width=\textwidth]{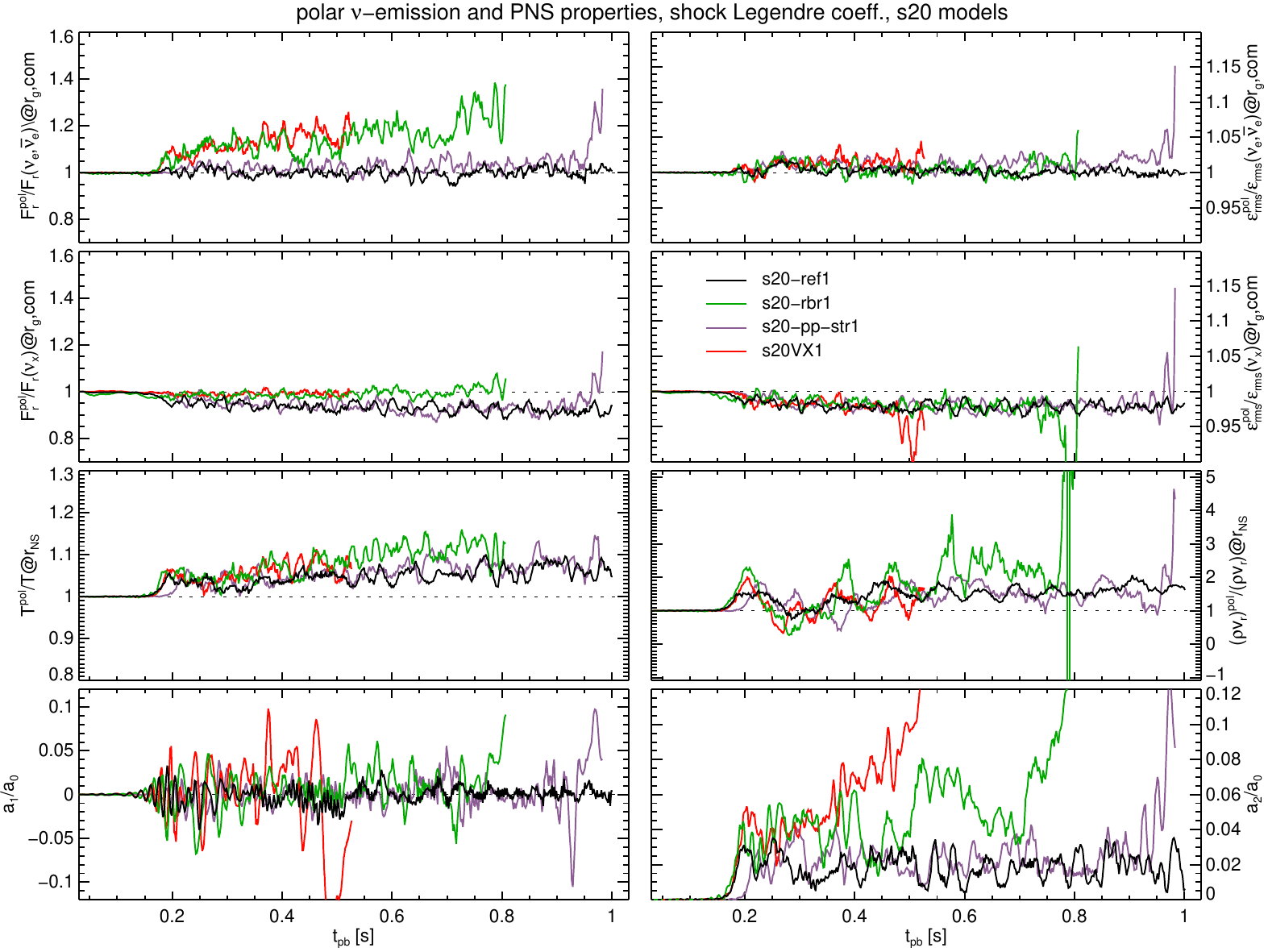}
 \caption{Impact of RbR+. Each panel in the first three rows shows for a given quantity at a given radius the value of this quantity averaged over the two cones with $\theta<\pi/6$ and $\theta>5\pi/6$, normalized to the average of this quantity over the full sphere. The first row shows this pole-to-sphere ratio for the energy-integrated radial fluxes (left panel) and rms-energies (right panel) summed over both species $\nu_e$ and $\bar\nu_e$ and measured at the gain radius in the comoving frame. The second row shows the same quantities but just for the $\nu_x$ neutrinos. The third row shows the pole-to-sphere ratio for the temperature (left panel) and radial mass-flux density (right panel) at the PNS surface (defined by the radius at which $\rho=10^{11}\,$g\,cm$^{-3}$). The bottom row shows the Legendre coefficients of the dipole (left panel) and quadrupole (right panel) deformation moments of the shock surface normalized to the monopole moments. All curves are smoothed using running averages of 10\,ms, except for the mass fluxes, where 30\,ms were used.}
\label{fig:rbrcomp_s20}
\end{figure*}

\begin{figure*}
  \includegraphics[width=\textwidth]{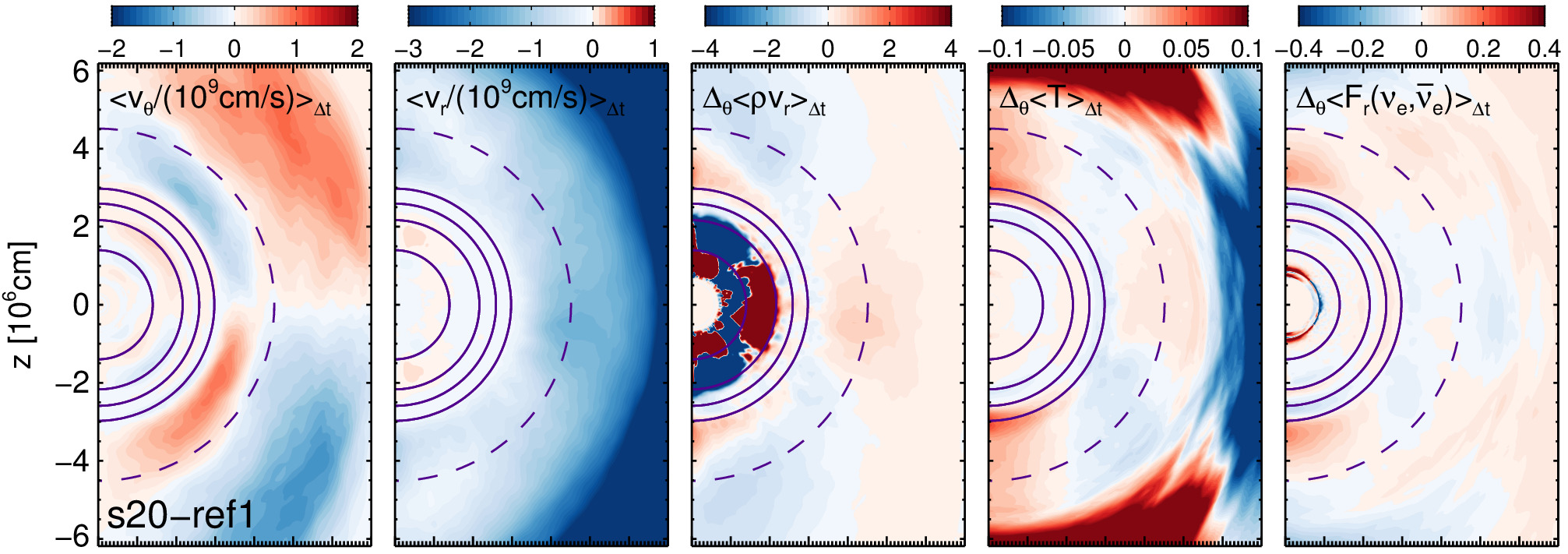}
  \includegraphics[width=\textwidth]{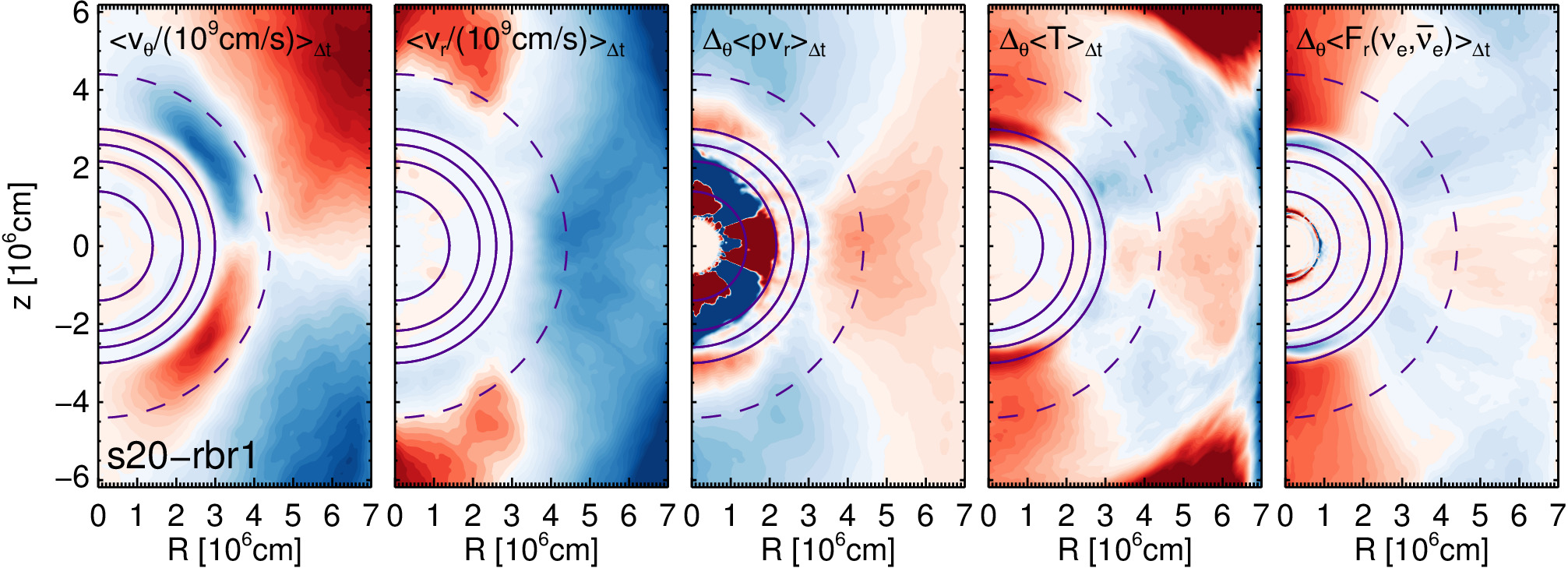}
  \caption{Color maps of time averages between $t_{\mathrm{pb}}=0.4\,$s and 0.6\,s of the lateral velocity, $v_\theta$, radial velocity, $v_r$, and angular variations of the radial mass-flux density, $\rho v_r$, temperature, $T$, and radial, energy-integrated electron-type neutrino fluxes, $F_{r,\nu_e}+F_{r,\bar\nu_e}$, for model s20-ref1 and its RbR+ counterpart s20-rbr1. Angular variations of time-averaged quantities are computed as local relative differences from angle averages. Values outside of the color-map ranges are clipped. The four solid lines in each panel denote iso-density surfaces for time-averaged densities $\langle\rho\rangle_{\Delta t}=10^{11,12,13,14}$\,g\,cm$^{-3}$, and the dashed line indicates the time-averaged gain radius. For both models the time-averaged flow exhibits features characteristic of SASI sloshing: During each sloshing cycle material at large radii, $r\ga 40-50\,$km, moves away from the poles, falls down towards the PNS surface near the equator, whereupon material at lower radii, $r\la 40-50\,$km, is pushed towards the poles of the opposite hemisphere. The red polar regions in the panels for $\Delta_\theta\langle \rho v_r\rangle_{\Delta t}$ at radii $r\approx 30-40\,$km indicate local (negative) extrema of the mass-flux density. These extrema are coincident with temperature maxima and locally enhanced neutrino emission. Switching on RbR+ in the s20 models leads to stronger SASI sloshing, probably as consequence of more collimated neutrino emission and hence more efficient heating-feedback at the poles. The less energetic SASI in model s20-ref1 also explains the lack of polar regions where the time-averaged velocities are positive, while such (red-colored) regions in $\langle v_r\rangle_{\Delta t}$ are visible for model s20-rbr1 at radii $r\ga 40-45\,$km.}
\label{fig:timeavg_s20}
\end{figure*}

\begin{figure*}
\centering
\includegraphics[width=\textwidth]{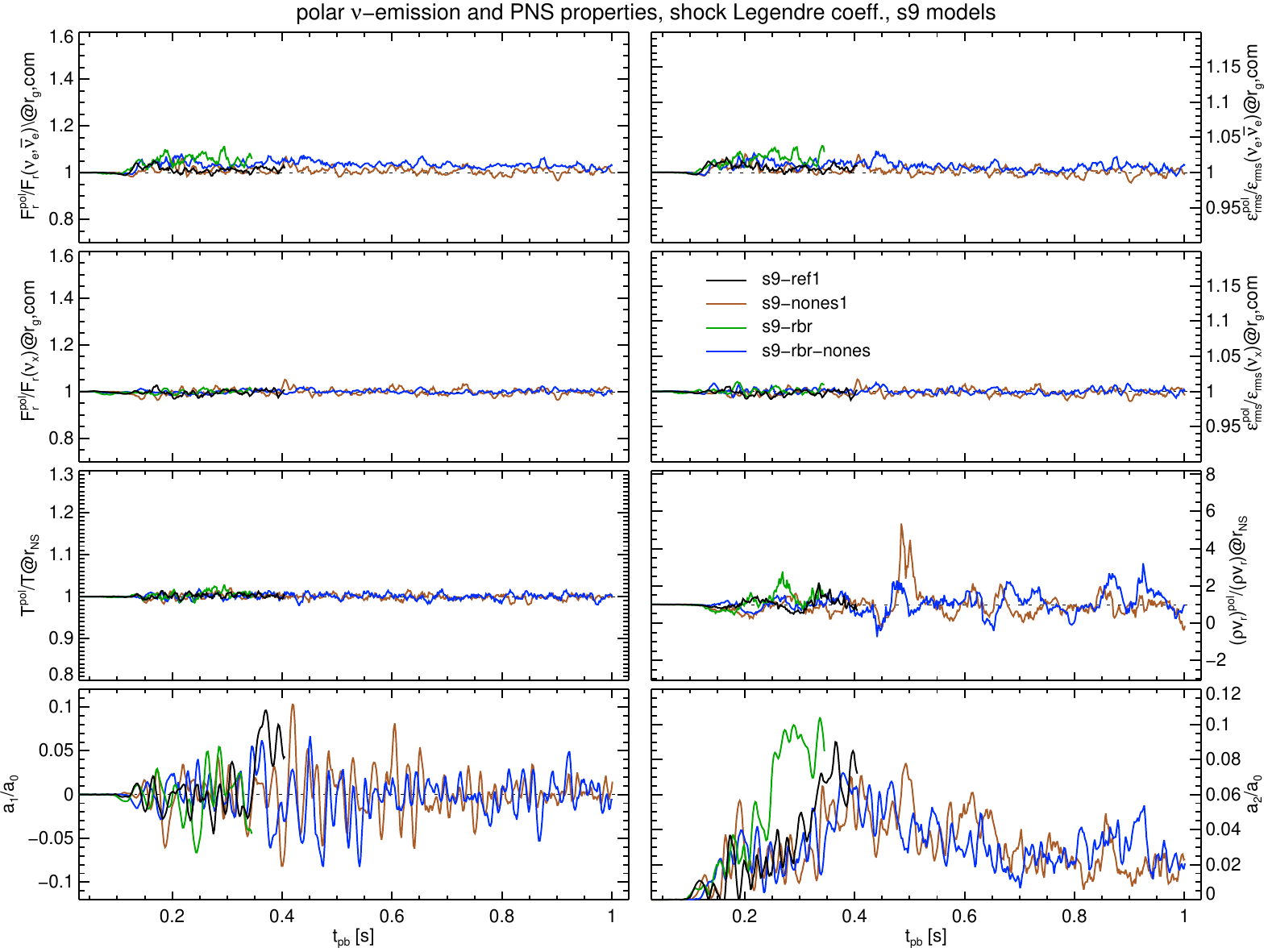}
 \caption{Same as Fig.~\ref{fig:rbrcomp_s20} but for models using the s9 progenitor.}
\label{fig:rbrcomp_s9}
\end{figure*}	

\begin{figure*}
  \centering
 \includegraphics[width=\textwidth]{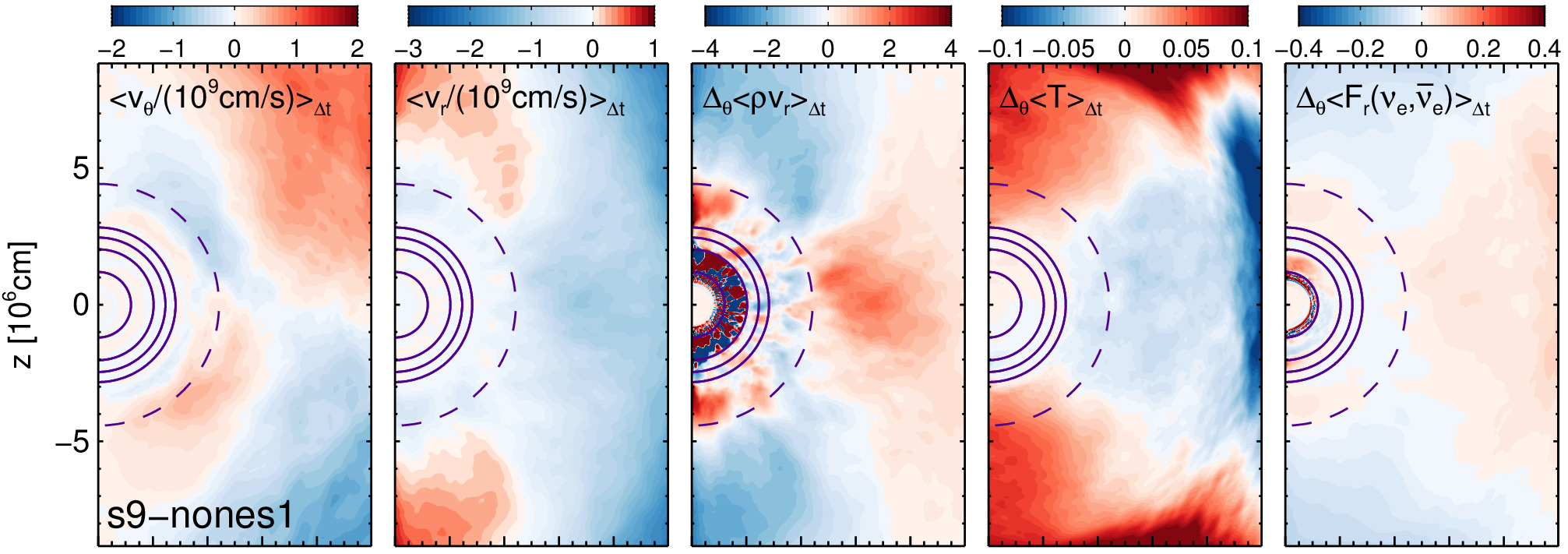}
 \includegraphics[width=\textwidth]{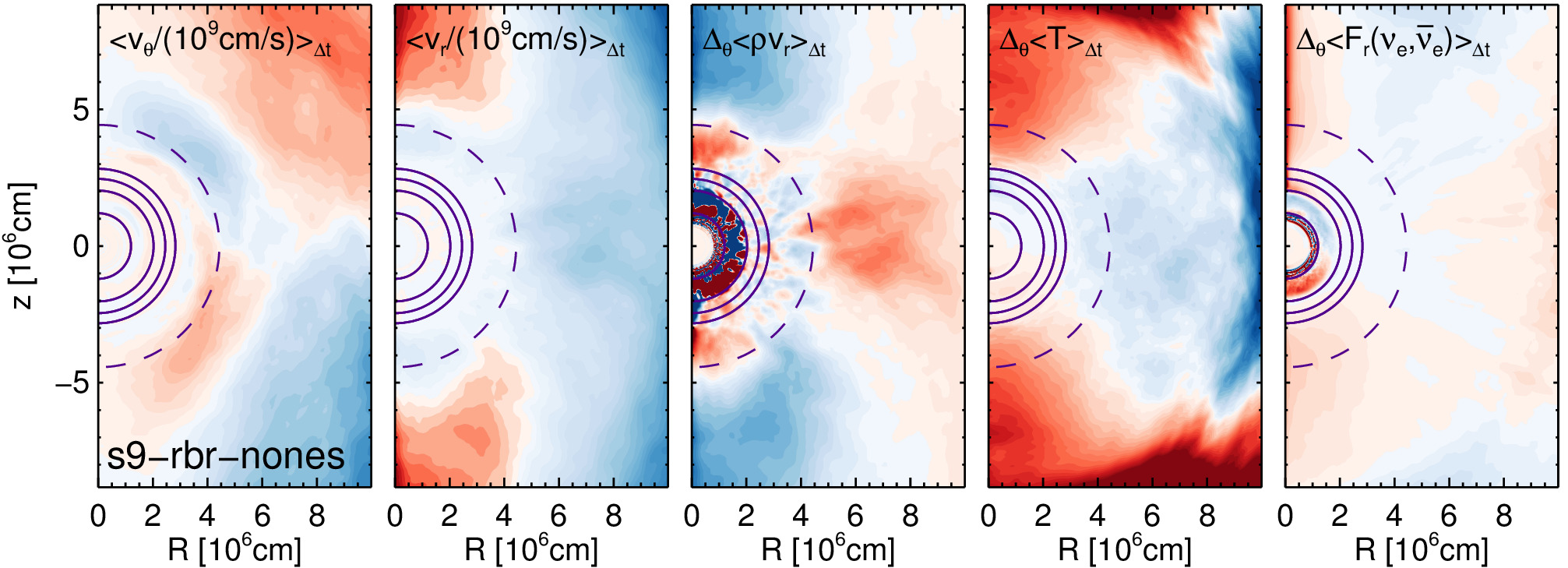}
 \caption{Same as Fig.~\ref{fig:timeavg_s20} but for model s9-nones1 and its RbR+ counterpart s9-rbr-nones. Note that while the color ranges are identical the spatial domain is slightly extended compared to Fig.~\ref{fig:timeavg_s20}. The time-averaged flow is mainly characterized by two large-scale convection cells operating quasi-independently in both hemispheres. On average, hot material rises at the poles and cold material is accreted close to the equator. Compared to the case of the $20\,\Msol$ models (cf. Fig.13),  accumulation of matter at the poles of the PNS is less efficient and the temperature- and emission-asymmetries at the PNS surface are correspondingly smaller. Overall, the results exhibit amazing agreement between the calculations performed with and without RbR+. }
 \label{fig:timeavg_s9}
\end{figure*}

\subsection{Ray-by-ray-plus approximation}\label{sec:ray-ray-plus}

In this section we compare the impact of using the RbR+ approximation on the PNS convection and neutrino emission as well as on the explosion dynamics.

\subsubsection{PNS convection and neutrino emission}\label{sec:pns-convection}

We start by addressing the question if using the RbR+ approximation might have a significant impact on the PNS convection and, in turn, on the neutrino emission from the deleptonizing PNS. Figure~\ref{fig:pnscomp} compares the angle-averaged, radial profiles of (comoving-frame) luminosities, $4\pi r^2 \int F_\nu \dd\eps$ for all evolved neutrinos species, $\nu\in\{\nu_e,\bar\nu_e,\nu_x\}$, as well as the radial profiles of the convective lateral kinetic energies, $\rho v_\theta^2/2$, at two representative times for models s20-ref and s20-rbr. The convectively unstable region is characterized by large values of $\rho v_\theta^2/2$. The neutrino luminosities -- all the way from inside the PNS up to the saturation (i.e. gain) radius -- as well as the convective energies are very similar for both models (apart from small temporal fluctuations) even at late post-bounce times. Moreover, as can be seen in Figs.~\ref{fig:timplot2d1} and~\ref{fig:timplot2d2} showing functions of time for the s20 models and Fig.~\ref{fig:timplot2d_s9} for the s9 models, also the mean and rms-energies of emitted neutrinos are barely affected. These results suggest that RbR+ has a negligible impact on PNS convection and on the angle-averaged neutrino emission, at least much less than, e.g., neglecting velocity-dependent terms in the transport (cf. Sec.~\ref{sec:velocity-terms}). This result is not too astonishing considering that neutrinos deep inside the PNS are strongly coupled to the medium such that, first, the lateral advection fluxes included in the RbR+ approximation strongly dominate the lateral diffusion fluxes, and second, lateral neutrino-momentum transfer to the fluid is well described by using pressure derivatives associated with assumed isotropic neutrino distributions, as applied in the RbR+ treatment \citep{Buras2006}.

\subsubsection{Impact on explosion behavior of s20 models}\label{sec:s20-models}

While the reference model, s20-ref, is not showing any indication of shock runaway during $\sim 1\,$s of post-bounce evolution, the corresponding RbR+ model, s20-rbr (which was compared against the corresponding \textsc{Vertex} model in Sec.~\ref{sec:comp-with-vert-1}), does explode, though rather late and with a substantial scatter in the time of shock runaway. Likewise, switching to the simplified pair-process treatment, which itself promotes explodability, yields late explosions without (model s20-pp) and early explosions with (s20-rbr-pp) the RbR+ treatment. In our set of s20 models, the net effect of RbR+ regarding the explodability is comparable to using the pair-process simplification together with either the strangeness (s20-pp-str) or the many-body correction (s20-pp-mb) but smaller than using all three of those together (s20-pp-str-mb).

In order to understand what pushes the RbR+ models closer to criticality, we consider Fig.~\ref{fig:rbrcomp_s20}, which shows for several models the excess/shortage of neutrino fluxes, rms-energies, temperatures, and mass fluxes close to the poles (i.e. averaged over the surfaces of two opposite cones defined by $\theta<\pi/6$ and $\theta>5\pi/6$) with respect to their averages over the full sphere, as well as the normalized Legendre coefficients for dipole and quadrupole deformation of the shock surface. All exploding and non-exploding s20 models with RbR+ (including the ones not plotted in Fig.~\ref{fig:rbrcomp_s20}) show a sustained polar enhancement of temperatures, mass-fluxes (albeit with large temporal fluctuations) and electron-type neutrino fluxes. At the same time, the RbR+ models come along with higher shock-oscillation amplitudes and therefore more optimistic runaway conditions. 

Better insight can be gained from Fig.~\ref{fig:timeavg_s20}, where we compare for models s20-ref1 and s20-rbr1 color maps of several quantities averaged in time between  $t_{\mathrm{pb}}= 0.4\,$s and $0.6$\,s, namely the lateral velocity, $v_\theta$, radial velocity, $v_r$, and angular variations of the radial mass-flux density, $\rho v_r$, temperature, $T$, and radial, energy-integrated flux density of electron-type neutrinos, $F_{r,\nu_e}+F_{r,\bar\nu_e}$. The angular variations are computed for each time average $\langle X\rangle_{\Delta t}$ of quantity $X$ as
  \begin{align}\label{eq:timavg}
    \Delta_\theta \langle X\rangle_{\Delta t} \equiv \frac{\langle X\rangle_{\Delta t}-\langle  X\rangle_{\Delta t,\theta}}{\langle X\rangle_{\Delta t,\theta}} \, ,
\end{align}
  where $\langle X \rangle_{\Delta t,\theta}\equiv (1/2)\int_{-1}^{1} \langle X\rangle_{\Delta t}
  \,\dd\cos\theta$ denotes the angle average of $\langle X\rangle_{\Delta t}$ at a given radius. For both models, the color maps of $\langle v_\theta\rangle_{\Delta t}$, $\langle v_r\rangle_{\Delta t}$ and $\Delta_\theta\langle\rho v_r\rangle_{\Delta t}$ carry the imprint of a large-scale, low-order flow pattern, which in the present case is mainly the result of quasi-periodic SASI sloshing: At large radii, $r\ga 40-50\,$km, post-shock material preferentially expands and moves away from the symmetry axis while falling back down towards the PNS. This causes large amounts of gas to arrive at the cooling region in the vicinity of the equator (cf. $\Delta_\theta\langle \rho v_r\rangle>0$ near $R\sim 40-50\,$km in Fig.~\ref{fig:timeavg_s20}), while the sloshing motion is further carried on by material moving towards the correspondingly other hemisphere at radii $r\la 40-50\,$km close to the PNS surface (e.g. $\langle v_\theta\rangle_{\Delta t} >0$ for $z<0$). This portion of the flow converges near the poles and helps driving the next half-cycle of approximately cigar-shaped shock expansion. Importantly, the converging flow also results in a net accumulation of matter near the poles. As can be seen in the panels for $\Delta_\theta\langle \rho v_r\rangle_{\Delta t}$ by the red-colored polar regions around radii of $r\approx 30-40\,$km, this surplus of matter leads to (in absolute terms) increased  mass-flux densities down onto poles of the PNS surface (the latter being roughly coincident with the largest purple circle denoting the $\rho=10^{11}\,$g\,cm$^{-3}$ surface of the time-averaged configuration). These accretion hot spots produce enhanced temperatures at and high radial neutrino fluxes streaming away from the poles, consistent with the results of Fig.~\ref{fig:rbrcomp_s20}.

  While the basic flow pattern and the polar temperature- and neutrino-flux enhancements are qualitatively similar in both models, the RbR+ model exhibits important differences compared to the unconstrained model: Non-radial motions are faster, polar expansion flows encountered during each SASI cycle are more powerful and even lead to positive time averages of radial velocities (red regions for $r\ga 40-45\,$km in the $\langle v_r\rangle_{\Delta t}$ panel in Fig.~\ref{fig:timeavg_s20}), and the polar accretion hot spots are characterized by higher temperatures and stronger neutrino emission. Moreover, neutrino radiation clearly remains more collimated above the polar hot spots than in the model without RbR+, where the radial fluxes relative to their angle averages quickly decrease with increasing radius, probably as a result of high lateral neutrino fluxes pointing away from the hot spots.

Based on the results described above it seems reasonable to suspect that a positive-feedback mechanism may be at play between the originally purely fluid-dynamical advective-acoustic/advective cycle \citep[e.g.][]{Foglizzo2002a, Blondin2006, Scheck2008, Guilet2012a} of the SASI and accretion-induced neutrino heating: Linear sloshing modes that lead to accretion hot spots at the PNS poles may get amplified due to absorption of neutrinos originating from these hot spots while, in turn, the neutrino emission rates at these hot spots may get boosted with growing amplitude of the sloshing modes. The efficiency of this feedback mechanism is higher with RbR+ than without, apparently because neutrino fluxes stemming from the hot spots remain more collimated when reaching the gain layer (see right panels in Fig.~\ref{fig:timeavg_s20}), causing more energy to be pumped into the fastest, near-axis material that carries most of the kinetic energy.
  
If our picture is correct, we may additionally speculate that the rather significant impact of RbR+ in the present SASI-dominated models is fostered by the following circumstances: First, sloshing modes generate strong downflows always at the same location, namely near the poles, which facilitates the development of hot spots and creates locally enhanced neutrino fluxes that are always pointed towards the optimal direction for dynamical feedback. Second, sloshing modes cause downflows to be rather well synchronized with expansion flows, i.e. both take place on the same characteristic timescales (namely the advection timescale $\tau_{\mathrm{adv}}$) and downflows are always quasi-periodically succeeded by expansion flows.

\subsubsection{Impact on explosion behavior of s9 models and comparison with s20 models}\label{sec:s9-models}

We now consider the convection-dominated s9 models;
see Figs.~\ref{fig:timplot2d_s9} and~\ref{fig:rbrcomp_s9} for the corresponding time-dependent properties of selected models, as well as Fig.~\ref{fig:timeavg_s9} for color maps of time-averaged quantities for two s9 models without and with RbR+.
Here, using RbR+ advances the onset time of shock runaway only by $\sim 0.02-0.13\,$s in model s9-rbr compared to s9-ref (cp. Table~\ref{table_models}), which is a rather small effect considering that neglecting neutrino-electron scattering prohibits the runaway entirely, both with (s9-rbr-nones) and without (s9-nones) using RbR+. The impact of RbR+ is thus much less extreme than in the s20 models, where switching off RbR+ delays the explosion by a much longer time than switching off neutrino-electron scattering (cp. models s20-rbr-pp, s20-pp, and s20-rbr-pp-nones, respectively).

The reduced impact of RbR+ for the s9 models becomes particularly obvious when comparing the two longest evolving, non-exploding models s9-nones and s9-rbr-nones (cf. Figs.~\ref{fig:timplot2d_s9},~\ref{fig:rbrcomp_s9}): The polar neutrino emission is only marginally enhanced by RbR+, and the dipole and quadrupole shock-deformation modes, the polar temperatures and mass accretion rates at the PNS surface, as well as essentially all global properties in Fig.~\ref{fig:timplot2d_s9} are nearly identical for both models.

What could be the reason for the much smaller impact of RbR+ in the s9 models compared to the s20 models? At least to some degree the answer must be connected to the fundamental difference between the two fluid instabilities respectively at play:  The SASI is a global instability that triggers quasi-periodic, linear sloshing motions between the two hemispheres. Neutrino-driven convection, on the other hand, instigates buoyancy modes (i.e. high-entropy bubbles) on, at least initially, smaller spatial scales than SASI, corresponding to higher multipole orders \citep{Foglizzo2006a}, and in a spatio-temporally much more stochastic fashion. It therefore seems plausible to suspect that large-scale asymmetries (associated with fluid modes of low multipole order) are typically smaller for a convection-dominated flow and, recalling the results for the s20 models, that a feedback mechanism between polar hot spots and neutrino heating is less likely to be realized.

However, this rather qualitative argument alone may be unsatisfactory without having considered in more detail the actual flow pattern resulting in the saturated state (cf. Figs.~\ref{fig:rbrcomp_s9} and~\ref{fig:timeavg_s9}). For this to be properly understood we recall, first, that axisymmetric models exhibit a designated axis parallel to which bubbles tend to rise more readily than in any other direction, and second, that they are subject to an inverted turbulent energy cascade, i.e. energy is transported from small to large scales and not the other way around as in 3D \citep{Kraichnan1967a, Hanke2012a, Couch2013b}. Consistent with these properties we observe that also the s9 models develop powerful low-order modes: The coefficients of dipolar ($a_1/a_0$) and quadrupolar ($a_2/a_0$) shock deformation modes in Fig.~\ref{fig:rbrcomp_s9} reach values as high as or even higher than those of the s20 models (cp. Fig.~\ref{fig:rbrcomp_s20}). Moreover, the panels for $\langle v_\theta\rangle_{\Delta t}$, $\langle v_r\rangle_{\Delta t}$, and $\Delta_\theta\langle \rho v_r\rangle_{\Delta t}$ in Fig.~\ref{fig:timeavg_s9} indicate that the time-averaged flow is dominated by a single large-scale convective eddy in each hemisphere: Hot bubbles rise preferentially near the poles, expand sideways while cooling down, re-enter the cooling layer near the equator, replace previously ascending bubbles at the poles, and finally start to rise again.

We caution the reader, however, that the circumstance that some qualitative features of the \emph{time-averaged} flow appear to be similar in Figs.~\ref{fig:timeavg_s9} and~\ref{fig:timeavg_s20} does not imply similar dynamical behavior of both types of models, s9 and s20, respectively. On the contrary: In the convection-dominated s9 models the flow pattern is rather quasi-stationary and in each hemisphere largely decoupled from that in the correspondingly other hemisphere. In contrast, in the SASI-dominated s20 models the flow pattern is much more variable with time, because post-shock material quasi-periodically sloshes from one hemisphere to another.

Coming back to the question why the impact of RbR+ is reduced in the s9 models: The panels for $\Delta_\theta\langle \rho v_r\rangle_{\Delta t}$ and $\Delta_\theta\langle T\rangle_{\Delta t}$ in Fig.~\ref{fig:timeavg_s9} show that even though the s9 models do exhibit vigorous low-order (i.e. dipole and quadrupole) convective modes, the mass-accretion rate in this progenitor is so low that the convective flow does not accumulate enough matter near the poles of the PNS surface to lead to dynamically relevant emission hot spots.

Considering only two progenitor models here we can hardly make general statements concerning the impact of RbR+ in axisymmetry that hold for all convection-dominated and SASI-dominated models, led alone the cases where such a distinction is not possible. However, our results may be indicative of the tendency that a convection-induced flow pattern composed of two quasi-stationary eddies is generically not as efficient as SASI sloshing in producing -- around the poles or at any other location on the PNS surface -- a significant surplus of matter that could lead to a dynamically relevant emission asymmetry. Perhaps an additional systematic reason speaking against the possibility of a positive-feedback cycle in convection-dominated models may be the lack of synchronicity between down- and expansion flows, i.e. the circumstance that the rise time of convective bubbles (roughly given by $\sim \tau_{\mathrm{heat}}$) is too long compared to the advection timescale, $\tau_{\mathrm{adv}}$, for bubbles to experience efficient feedback from neutrino emission that is released by downflows on an advection timescale. In other words, the enhancement of polar neutrino emission triggered by some polar downflow declines too quickly for the next convective bubble to reach appreciable positive radial velocity. This argument could at least explain why for the two exploding models, s9-rbr and s9-ref, we actually do see an, although small, explosion-promoting impact of RbR+, because here $\tau_{\mathrm{heat}}$ is obviously closer to $\tau_{\mathrm{adv}}$ (cf. Fig.~\ref{fig:timplot2d_s9}), at least shortly before shock runaway sets in.

\subsection{Stochasticity and resolution dependence}

We have already found (see, e.g., Sec.~\ref{sec:comp-with-vert-1}) that the impact of stochasticity and temporal fluctuations is low regarding the emission related properties but can be substantial for the gain-layer related properties including the onset time of explosion. In this section we want to briefly consider the results provided by our models that allow one to identify some systematic tendencies. Subsequently, we address the question how well the obtained global features, including the scatter of shock-runaway times, are converged with respect to the numerical grid.

Comparing the time evolution of $\tau_{\mathrm{adv}}/\tau_{\mathrm{heat}}$ (Figs.~\ref{fig:timplot2d1},~\ref{fig:timplot2d2} and~\ref{fig:timplot2d_s9}) between various models reveals, first, that the amplitudes of temporal fluctuations are lower in the convection-dominated s9 models than in the SASI-dominated s20 models, and second, that at least for the s20 models these amplitudes grow with the proximity of each model to the runaway threshold. This tendency is consistent with the observed time interval, $\Delta t_{\mathrm{exp}}$, within which the runaway times are dispersed for models differing only in the initial perturbation pattern (see Table~\ref{table_models}): The scatter seems to be greater for the s20 models than for the s9 models, and for the s20 models the scatter is larger for more marginally exploding models (e.g. s20-rbr, s20-pp-str) than for more robustly exploding models (e.g. s20-rbr-pp). These tendencies are in qualitative agreement with \citet{Cardall2015a}, who examined stochasticity using a large number of simplified models and found larger dispersion of explosion times for SASI-dominated models than for convection-dominated models \citep[see also][]{Kazeroni2017a}.

The strong dependence of $\Delta t_{\mathrm{exp}}$ on the progenitor and the input physics might be one reason why previous studies report quite diverse values of $\Delta t_{\mathrm{exp}}$ \citep[e.g.][]{OConnor2018a,Summa2016a,Cardall2015a,Takiwaki2014a}. An additional reason might simply be that many studies, particularly the ones including computationally expensive neutrino transport such as ours, are forced to rely on rather poor statistics because they can only afford a small number of simulations.

Coming now back to the second question concerning numerical convergence: For the reference s20 model, s20-ref, and its counterpart including the RbR+ approximation, s20-rbr, we repeated the simulations with both increased and decreased resolutions in both radial and angular directions (see models ending with ``hires'', ``lores'', ``hi$\theta$'' and ``lo$\theta$'' in Table~\ref{table_models}), in some cases even multiple times with different initial perturbation patterns. We could not identify a systematic trend with varying the resolution, neither regarding the neutrino-emission properties nor the heating conditions in the gain layer nor the scatter in explosion times. In particular, for model s20-rbr the onset of explosion does not appear to be correlated with resolution in any direction: For all three angular resolutions with 80, 240, and 320 zones the scatter $\Delta t_{\mathrm{exp}}$ in the times of shock runaway remains comparably high, namely, $\Delta t_{\mathrm{exp}}= 0.33\,$s, 0.44\,s, and 0.43\,s, respectively. Since model s20-rbr is quite marginal concerning its tendency to explode and would therefore probably be quite sensitive to numerical resolution, this suggests (although does not prove) that, at least, features the runaway is sensitive to are numerically converged.

\subsection{Previous axisymmetric simulations of the s20 progenitor}

Here we want to collect some results concerning the explosion behavior of axisymmetric simulations performed by various other groups using the s20 progenitor. A detailed comparison is, however, out of the scope of this paper.

\citet{OConnor2018a} used M1 transport in cylindrical coordinates, a similar neutrino setup as applied here for model s20-pp but additionally neglecting neutrino-electron scattering, and employing the LS220 EOS \citep{Lattimer1991}. In their models shock expansion sets in around 700\,ms post bounce, while our model s20-pp explodes later and would probably not explode at all without neutrino-electron scattering. However, the LS220 EOS used in \citet{OConnor2018a} is slightly softer than the SFHo EOS used here and might lead to earlier explosion times.

Formally the same setup and a similar M1 scheme as in \citet{OConnor2018a}, although in spherical coordinates, was used by \citet{Skinner2016}. We were unable to ascertain how \citet{Skinner2016} treated pair processes, e.g. if they ignored annihilation for electron-type neutrinos and used an LTE assumption for $\nu_x$ pair-annihilation partners as in \citet{OConnor2018a} and as in our ``pp''-models, or if they treated pair processes like we did in all other models. In the latter case, the lack of an explosion both with and without RbR+ would be consistent with our models, while in the former case it could mean that their models explode less readily than ours.

\citet{Kotake2018a} employed an IDSA scheme in the RbR+ mode, the LS220 EOS, and a variety of different neutrino interactions, while energy-bin coupling reactions (neutrino-electron scattering as well as pair processes) are included up to 0th angular order of the interaction kernels. Their model G1 is most similar to our model s20-rbr-pp. While their model G1 does not explode until the simulation was stopped at 600\,ms, our model explodes around 350\,ms. However, the comparison is not conclusive because model G1 ignores weak magnetism and recoil corrections after \citet{Horowitz2002}.

Finally, \citet{Bruenn2016a} and \citet{Summa2016a}, though using mutually different transport solvers, both employed RbR+ and an advanced (compared to the one used here) set of neutrino interactions and the LS220 EOS. In \citet{Bruenn2016a} this model (as well as all others in that study) started shock expansion around 150\,ms, while in \citet{Summa2016a} this happened only around 300-350\,ms.

Although the diversity of these results is quite considerable, this is not too astonishing in view of the fact that few-percent variations in a single neutrino interaction channel may already shift the time of shock runaway, $t_{\mathrm{exp}}$, by several hundred milliseconds.

\section{Summary and conclusions}\label{sec:summary-conclusions}

In this study we used 1D (spherically symmetric) and 2D (axisymmetric) models to compare the relatively new \textsc{Aenus-Alcar} code \citep{Just2015b}, which incorporates the fully multidimensional M1 approximation for neutrino transport, against the well-established \textsc{Prometheus-Vertex} code \citep{Rampp2002,Buras2006}, which employs an accurate Boltzmann solver restricted to the ray-by-ray+ (RbR+) approximation that neglects non-radial neutrino flux components. We compared with \textsc{Vertex} by mimicking the RbR+ approximation in \textsc{Alcar} and we tested the RbR+ approximation by comparing to the fully multidimensional version of \textsc{Alcar}. Moreover, we investigated the impact of other modeling variations in the neutrino transport that are frequently used by CCSN modelers, namely neglecting inelastic neutrino-electron scattering, simplifying pair-processes by assuming target neutrinos to be in isotropic equilibrium, applying strangeness and many-body corrections to neutrino-nucleon scattering, and ignoring velocity-dependent as well as gravitational redshift terms in the transport equations. Starting from spherically symmetric and non-rotating progenitor models, asymmetries are not expected to develop during the core-collapse phase, which therefore can be simulated by using computationally less demanding 1D calculations. Moreover, since the impact of some of our modeling variations on the (one-dimensional) core collapse has already been studied in detail before \citep[e.g.][]{Lentz2012a, Lentz2012b}, we only focus here on the ramifications of these variations on the post-bounce evolution, by initializing all post-bounce models with the same collapse model. Finally, in order to obtain information about the degree of stochasticity in our comparison study we repeated some of our simulations several times starting with different initial random perturbation patterns (but the same perturbation amplitudes). With respect to the questions raised in the introduction, we obtained the following results:

\begin{enumerate}[wide,labelwidth=!,labelindent=0pt,label=\arabic*.,leftmargin=3mm]
\item In 1D the agreement between \textsc{Alcar} and \textsc{Vertex} is found to be excellent concerning nearly all features. The only noteworthy differences are a more energetic neutrino burst, slightly ($\la 5\,$\%) higher luminosities, and a hotter and (by $\approx 1\,$km) more compact proto-neutron star (PNS) in \textsc{Alcar}.

\item In the two examined 2D models of the SASI-dominated s20 progenitor with and without neutrino-electron scattering, the agreement found between \textsc{Alcar} in the RbR+ mode and \textsc{Vertex} remains very good concerning the neutrino emission, the PNS contraction, the heating conditions, and the explosion times. The two last mentioned features are subject to substantial stochastic scatter, the degree of which is, consistently in both codes, stronger for the exploding models with neutrino-electron scattering than for the non-exploding models without. Similar to the 1D case, the PNS radius in the \textsc{Alcar} models is again smaller throughout by $\sim 1\,$km compared to the \textsc{Vertex} models. Although PNS convection has almost the same impact on the PNS radius and neutrino luminosities with \textsc{Alcar} and \textsc{Vertex}, the associated kinetic energies are higher by a factor of a few in \textsc{Alcar} than in \textsc{Vertex}. The origin of this discrepancy is not related to the RbR+ approximation and needs to be found in future work.

\item When comparing \textsc{Alcar} models with and without the RbR+ approximation, we could not observe any significant sensitivity of PNS convection and of the (angle-averaged) neutrino emission on the use of RbR+. Concerning the explosion behavior, we find that the RbR+ approximation only becomes noticeable once long-lived (polar) hot spots appear, above which material is heated more efficiently than without RbR+. In the investigated models hot spots are formed as a result of large-scale, low-order fluid modes that accumulate matter near the poles of the neutrinosphere. We observe clear explosion-promoting consequences for the SASI-dominated s20 models, but only a weak impact for the convection-dominated s9 models. We interpret this as a consequence of the tendency that linear sloshing modes may be more efficient than convection-driven modes in creating long-lasting accretion hot spots on the proto-neutron star surface. However, even in the SASI-dominated models the net impact of RbR+ remains manageable and quantitatively comparable to typical modeling variations in the microphysics sector (see next item).
  
\item Simplifying pair processes, including neutrino-electron scattering, as well as adopting the strangeness and many-body corrections during the post-bounce evolution all have, roughly in this order of relevance, a significant explosion-facilitating impact in our 2D models, while in 1D these modifications result in rather small changes of the shock radius. The net effect on the explosion times of combining the two considered opacity simplifications (regarding pair processes and neutrino-electron scattering) is smaller than that of using either simplification individually (cp. models s20-rbr, s20-rbr-nones, s20-rbr-pp, s20-rbr-pp-nones in Table~\ref{table_models}). The timescale ratio, $\tau_{\mathrm{adv}}/\tau_{\mathrm{heat}}$, in 1D predicts on a qualitative level remarkably well the impact of each modeling variation for the 2D models. In contrast, the shock radius can be shifted both to lower (``str'' and ``mb'' models) or higher (``pp'' models) values in 1D for microphysics variations that promote an explosion in 2D. This suggests that $\tau_{\mathrm{adv}}/\tau_{\mathrm{heat}}$ is a more powerful diagnostic quantity than the shock trajectory when estimating the impact of modeling variations using computationally less demanding 1D models.

\item Ignoring velocity-dependent terms and gravitational redshift in the transport equations during the post-bounce evolution reduces the neutrino fluxes noticeably and the rms-energies barely as measured in the comoving frame of infalling material in the gain region. In 2D these explosion-hampering features can, however, be (over-)compensated, because PNS contraction is accelerated owing to less vigorous PNS convection. The net effect on the explosion is case dependent and becomes more pessimistic for more massive PNSs, which imply higher velocities and stronger redshift: For an s20 model that originally explodes early when the PNS is still less massive, the explosion sets in earlier, while for an originally late exploding model the explosion lacks entirely. For the low-mass s9 models the aforementioned effects compensate and there is barely any visible impact on the shock trajectory. 

\item The conditions in the gain layer (e.g. the lateral kinetic energy or timescale ratio) and the explosion times can be subject to substantial random variations of several hundred milliseconds, in agreement with the findings  by \citet{Cardall2015a} based on simplified models. The amplitudes of temporal fluctuations and the scatter in the explosion times increase with the proximity to criticality for a given model, and they turned out to be much greater for the SASI-dominated s20 models than for the convection-dominated s9 models.
  
\end{enumerate}

The result that our systematic comparison reveals good agreement between \textsc{Alcar} and \textsc{Vertex} is reassuring with respect to the numerical implementation of the codes and employed input physics, and it may also be considered as mutual support of the basic viability of the approximations made in both schemes, namely M1 and RbR+, at least for problems and setups similar to those considered in this study.
    
  The tests of RbR+ confirm the previous suspicion \citep{Dolence2015a, Sumiyoshi2015a,Skinner2016} that RbR+ can facilitate explosions, but they do so only partly, because our convection-dominated models are only marginally affected and a corresponding comparison in the more realistic three-dimensional case has yet to be conducted. Also, one should keep in mind that our comparison test of RbR+ bears uncertainties that are connected to the still incompletely known accuracy of fully multidimensional M1. For instance, it might be possible that M1 over- or underestimates the lateral dilution of radiation emitted from polar hot spots, in which case the observed differences between the SASI-dominated s20 models would presumably be over- or underrated here, respectively.

Remarkably, significant effects of RbR+ only seem to enter the evolution by means of long-lived, low-order multipole modes of the flow. In contrast, RbR+ effects triggered by temporary, stochastic downflows seem to remain weak on dynamical timescales \citep[in agreement with the expectation of][]{Buras2006}, even though the instantaneous radiation field may exhibit significant local anisotropies with RbR+ \citep[as found by][]{Sumiyoshi2015a}. This suggests that in the more relevant 3D case the consequences of using RbR+ might be overall less dramatic than in 2D, because an artificial symmetry axis that fosters axis-parallel motions is absent. If confirmed, in turn, a reduced impact of RbR+ in 3D would speak in favor of existing 3D results obtained using the RbR+ approximation and would justify using RbR+ in the future.

Although the remaining results concerning the investigated simplifications are qualitatively consistent with previous studies emphasizing the importance of various aspects of neutrino transport \citep[e.g.][]{Buras2006, Muller2012b, Lentz2012a, Sumiyoshi2015a, OConnor2018a, Burrows2018a, Richers2017a, Kotake2018a}, the rather high sensitivity with respect to ostensibly small transport details combined with a considerable level of stochasticity is somewhat surprising. The following considerations are therefore not entirely new, but are certainly strengthened by this comparison study:
\begin{enumerate}[wide,labelwidth=!,labelindent=0pt,label=\alph*.,leftmargin=3mm]

\item Profound knowledge of neutrino interaction rates is imperative. Further exploration of corrections to commonly used cross sections \citep[e.g.][]{Horowitz2017a} and of new neutrino physics \citep[e.g.][]{Bollig2017a} will more than likely have a significant leverage on future CCSN models. Our results suggest that a noticeable impact on the time of shock runaway can potentially be expected once a certain correction (or approximation) is large enough to cause a shift of the shock radius of just a few percent in corresponding spherically symmetric models.

\item The inclusion of (special and general) relativistic effects in multidimensional CCSN simulations is recommended, not only for obtaining a more precise time of shock runaway but also in order to achieve more reliable results for the PNS contraction and the luminosities and spectral properties of emitted neutrinos.

\item Careful analysis is necessary when comparing two models in the literature, particularly if these models use different approximations made in the neutrino sector and those differences are not rigorously tested and documented. In fact, such a comparison may turn out to be extremely difficult and quite amenable to premature conclusions. For instance, large differences in explosion times might be caused just by a different treatment of a certain neutrino interaction channel and may therefore not necessarily reflect more serious code differences. Likewise, good agreement between explosion times may not necessarily imply that all the individual differences are small; it could just happen that some differences accidentally cancel each other and the net impact is small. Moreover, due to their strongly non-linear nature, certain neutrino reactions may become more or less relevant when coupled with other reactions \citep[see, e.g.,][]{Lentz2012a}.

\item Tightly related to the previous item, the uncertainties due to stochasticity need to be estimated and accounted for whenever drawing conclusions regarding mutually different physics ingredients of two simulations in 2D. Obviously, performing expensive stochasticity tests is barely feasible (and might not be as important) for the currently most detailed 3D models. However, with the number of available models growing and the wealth of experience increasing also in 3D applications, more light will be shed on the quantitative impact of stochasticity and its dependence on the progenitor model and on other conditions.

\end{enumerate}  

Nonetheless, while our results certainly suggest to exercise caution when interpreting the runaway times of CCSN models using different neutrino treatments, we should also be aware that the level of sensitivity might not always be as dramatic as seen in the present study. The progenitor models chosen here, in particular the s20 model, might be more marginal than other, possibly more representative models. Moreover, once the set of physics ingredients leads to more robustly exploding models, details may tend to matter less, which means that errors introduced by some approximations (such as RbR+, M1, or neglecting neutrino-electron scattering) would become less significant.

A few final remarks are in order concerning the approximate M1 closure. Although \textsc{Alcar} with RbR+ and \textsc{Vertex} show very good agreement, at this point we are unable to quantify the error introduced in dynamical simulations by the fully multidimensional M1 closure \citep[see, however,][for a comparison of M1 with different Boltzmann solvers based on stationary configuration snapshots]{Richers2017a}. At least the error cannot be dramatically large in all cases of models and progenitors, since otherwise we would have seen stronger differences between the s9 models with and without RbR+. Nevertheless, it is known \citep[e.g.][]{Pons2000} that M1 can lead to unphysical features in the case of crossing radiation beams, and comparisons with more sophisticated transfer schemes in the case of black-hole torus systems \citep[Appendix of][]{Just2015a} or differentially rotating neutron stars \citep{Foucart2018a} as remnants of neutron-star mergers suggest that the accuracy of M1  systematically decreases with increasing geometric complexity of the source. The M1 scheme might therefore be somewhat more accurate for ordinary CCSNe than for more exotic ones with high rotation rates and/or strong magnetization, for which significant neutrino-emission asymmetries appear. Still, even for those cases the M1 method is currently one of the most attractive schemes regarding accuracy and computational efficiency, considering that for highly aspherical geometries the RbR+ approximation is not advisable and given the scarcity of alternative, computationally feasible multidimensional transport methods. In any case, comparisons with more accurate methods based on dynamical simulations are needed in order to identify the relevant shortcomings of M1, and to assess for which physics questions it will be necessary to employ more involved and expensive Boltzmann solvers.

More comparisons such as the present one are needed to understand differences in existing multidimensional CCSN results obtained worldwide. These comparisons should include more codes using different transport approximations (e.g. FLD, IDSA), as well as discretization schemes for the hydro (e.g. cartesian or cylindrical or spherical polar grid) and the transport (e.g. tangent-ray, discrete ordinate, Monte Carlo). Our results suggest that convection-dominated progenitors, which seem to come with an overall smaller level of stochasticity, might be more suited for such studies than SASI-dominated progenitors.

\section*{Acknowledgments}
OJ is indebted to Thomas Ertl, J\'{e}r\^{o}me Guilet, Raphael Hix, Remi Kazeromi, Tobias Melson, Tomoya Takiwaki, Kohsuke Sumiyoshi, and Alexander Summa for helpful discussions. Moreover, we are grateful to Evan O'Connor and Almudena Arcones whose initiation of a community wide comparison project inspired parts of this work. The work of OJ, THJ, RB, and RG was supported by the European Research Council through grant ERC AdG 341157-COCO2CASA and the Cluster of Excellence “Universe” EXC 153, moreover for OJ and THJ by the Max-Planck–Princeton Center for Plasma Physics (MPPC).
MO acknowledges support from the European Research Council (grant CAMAP-259276) and from the Spanish Ministry of Economy and Finance and the Valencian Community grants under grants AYA2015-66899-C2-1-P and PROMETEOII/2014-069. OJ and SN are supported by the Special Postdoctoral Researchers (SPDR) program and the iTHEMS cluster at RIKEN. We acknowledge computational support by the Max Planck Computing and Data Facility (MPCDF) and the HOKUSAI supercomputer at RIKEN.



\appendix

\section{Treatment of source terms in \textsc{Alcar}}\label{sec:comp-source-terms}

In \textsc{Alcar} the overall time integration of the neutrino moments is performed together with the hydrodynamic variables using a 2nd-order Runge-Kutta scheme, in which all terms except some neutrino-interaction rates are treated explicitly in time. For the motivation of the overall integration method we refer the reader to \citet{Just2015b}. Here we describe the specific treatment of neutrino-interaction rates employed for the simulations in this paper.

Ignoring for now the Runge-Kutta sub-stepping, we advance the hydrodynamic equations, Eqs.~(\ref{eq:hydevo}), and neutrino-transport equations, Eqs.~(\ref{eq:momeq}), from an old time step, $t^\n$, to a new (partial) time step, $t^{\n+1} = t^\n+\Delta t$, as follows:
\begin{align}\label{eq:srcupdate}
  \mathbf{U}^{\n+1}  = \quad \mathbf{U}^{\n}  + \Delta t\left(\delta_t\mathbf{U}\right)^{\n,\n+1}_{\mathrm{src}}  - \Delta t\left(\delta_t\mathbf{U}\right)^\n_{\mathrm{other}}
\end{align}
where $\mathbf{U}^{\n/\n+1}=(\rho,\rho Y_e,\rho v^i,e_{\mathrm{t}},E_{\nu,q},F^i_{\nu,q})^{\n/\n+1}$ (with $\nu=\nu_e,\bar\nu_e,\nu_x$ and $q=1,\ldots,N_\eps$ denoting the neutrino species and energy group, respectively) is the vector of hydrodynamic and neutrino-transport variables at the old/new time step, $\left(\delta_t\mathbf{U}\right)^{\n,\n+1}_{\mathrm{src}}$ are the neutrino-source terms, which may depend on both old and new variables, and $\left(\delta_t\mathbf{U}\right)^\n_{\mathrm{other}}$ represents all remaining terms, which depend only on the old variables. Now if in Eq.~(\ref{eq:srcupdate}) all neutrino-interaction terms for the three species, four moments, and $N_\eps$ energy groups were to be treated fully implicitly, at each spatial grid point a non-linear system of equations of rank $3\times 4 \times N_\eps+2$ (the additional two variables are $Y_e$ and $T$) would need to be solved, which involves the inversion of a non-sparse Jacobian possibly multiple times per integration step. Given that the time steps used in \textsc{Alcar} are rather short (because of the explicit integration of the non-local divergence terms) and the total number of integration steps is therefore large, the computational cost would in this case soon become very large. This is why we avoid an implicit integration of energy-bin coupling source terms wherever justified, in the manner that is described below.

For all emission/absorption as well as iso-energetic scattering processes the source terms entering the moment equations, Eqs.~(\ref{eq:momeq}), are given by:
\begin{subequations}\label{eq:srcemab}
\begin{align}
   C^{(0)}_{\mathrm{e/a/s}} & = c\kappa^\star_{\mathrm{a}}(E^{\mathrm{eq}}-E)  \, , \\
   C^{(1),i}_{\mathrm{e/a/s}} & = -c(\kappa^\star_{\mathrm{a}}+\kappa_{\mathrm{s}})F^i \, ,
\end{align}
\end{subequations}
where $E^{\mathrm{eq}}$ is the equilibrium energy density corresponding to a Fermi-Dirac distribution, $\kappa^\star_{\mathrm{a}}$ is the sum of all absorption opacities corrected for stimulated absorption, and $\kappa_{\mathrm{s}}$ is the sum of opacities for iso-energetic scattering. We follow \citet{Rampp2002} for the computation of these opacities, except that we additionally account for weak magnetism and nucleon recoil \citep[][]{Horowitz2002} and, in selected models (cf. Sec.~\ref{sec:neutr-inter-invest}), for strangeness corrections and many-body corrections in a way described in \citep{Horowitz2017a}. For neutral-current reactions of $\nu_x$ neutrinos (representing the four heavy-lepton neutrinos) we use as effective opacity the arithmetic average of the opacity of a heavy-lepton neutrino and that of its antiparticle. In computing the source terms, Eqs.~(\ref{eq:srcemab}), the neutrino energy- and flux-densities are treated implicitly and the opacities explicitly. The equilibrium energy density, $E^{\mathrm{eq}}$, is usually treated explicitly, i.e. using old values of $Y_e$ and $T$. Occasionally, however, in regions with strong neutrino-matter coupling and short fluid-dynamical timescales the difference between equilibrium energies at the old and new time step becomes so large that numerical oscillations would result for an explicit treatment of $E^{\mathrm{eq}}$. In these (rare) cases, which we detect using a criterion based on the density and the relative change of gas energy due to neutrino interactions, we perform an additional intermediate step to obtain improved values of $Y_e$ and $T$ (and therefore $E^{\mathrm{eq}}$). In this step we solve Eqs.~(\ref{eq:srcupdate}) for $Y_e$ and $T$ using implicit equilibrium energies and under the simplifying assumption that the neutrino fluxes vanish.

Inelastic scattering of (all types of) neutrinos off electrons and positrons is implemented as in \citet{Yueh1977, Bruenn1985, Rampp2002}, i.e. using a Legendre expansion in the scattering angle of neutrinos up to 1st order. For the multidimensional generalization of the formalism of \citet{Yueh1977, Bruenn1985, Rampp2002} (who assumed spherical symmetry and therefore vanishing non-radial flux-vector components and non-diagonal Eddington-tensor components) we use the expressions given in \citet{Cernohorsky1994a}. We obtain the Legendre coefficients of up to 1st order for each initial-state and final-state neutrino energy and each neutrino species by table interpolation. The scattering rates are computed explicitly, i.e. using Legendre coefficients and neutrino moments from the old time step. In order for the explicit integration to remain stable, the rates for inelastic neutrino-lepton scattering are reduced at high densities (where these rates are subdominant compared to charged-current reactions for conditions considered in this paper) by a factor of $\max\{1,(\rho/(5\times 10^{12}\,\mathrm{g\,cm}^{-3})^{3/2}\}$ \citep[following the suggestion by][]{OConnor2015a}.

Finally, for pair processes we again follow \citet{Rampp2002} as closely as possible\footnote{We note the following misprints in the Appendix of \citet{Rampp2002}, which are correctly accounted for in both simulation codes, \textsc{Alcar} and \textsc{Vertex}: In Eq.(A.20), i.e. the multipole expansion of the 1st-moment source term for pair-processes, the factor $(2l+1)$ must be replaced by 1, in Eq.(A.39) the quantity erroneously denoted as production coefficient, $\phi^\mathrm{p}_0$, is actually the annihilation coefficient, $\phi^\mathrm{a}_0$, and in Eq.(A.41) the factor $(\hbar c)^3$ must be replaced by 1.}, where electron-positron annihilation is implemented based on \citet{Pons1998} and nucleon-nucleon bremsstrahlung is implemented based on \citet{Hannestad1998}. We expand the interaction kernel in $\cos\omega\equiv \vecn\cdot\bar\vecn$, where $\vecn$ and $\bar\vecn$ are the propagation unit vectors of the considered neutrino and its annihilation partner, respectively. The resulting expansion of the pair-process source terms for the moments $E$ and $F^i$ of a given neutrino species at energy $\eps$ reads (up to 2nd order in $\cos\omega$):
\begin{subequations}\label{eq:srcpair}
\begin{align}
  C^{(0)}_{\mathrm{pair}}   = & \frac{8\pi^2 c \eps^3}{(h c)^3} \int_0^\infty\dd\bar\eps\,\bar\eps^2
  \left[ \phi^\mathrm{p}_0(1-M_0-\overline{M}_0) + \phi_0^\mathrm{a}M_0 \overline{M}_0
  \right. \nonumber\\
  & + \left. 3 \phi_1^\mathrm{a}M_{1,j} \overline{M}_1^j + \frac{5}{2}\phi^\mathrm{a}_2(3M_{2,jk}\overline{M}_2^{jk} - M_0 \overline{M}_0) \right] \, , \\
  C^{(1),i}_{\mathrm{pair}}  = & \frac{8\pi^2 c^2 \eps^3}{(h c)^3} \int_0^\infty\dd\bar\eps\,\bar\eps^2
  \left[ -\phi^\mathrm{p}_0 M_1^i - \phi^\mathrm{p}_1\overline{M}_1^i
  \right.\nonumber\\
  & + \left. \phi_0^\mathrm{a}M_1^i \overline{M}_0 + 3 \phi_1^\mathrm{a}M_{2}^{ij}\overline{M}_{1,j} \right.\nonumber\\
  & + \left.\frac{5}{2}\phi^\mathrm{a}_2(3 M^{3,ijk}\overline{M}_{2,jk} - M_1^i \overline{M}_0)  \right]
\end{align}
\end{subequations}
where
\begin{align}
  \{ M_0, M_1^i, M_2^{ij}, M_3^{ijk}\} \equiv \frac{1}{4\pi}\int \dd\Omega \mathcal{F}
  \{1,n^i,n^{ij},n^{ijk}\}
\end{align}
are angular moments of the neutrino distribution function, $\mathcal{F}$, and
overlined symbols denote the corresponding quantities of the annihilation partner.
The coefficients $\phi_{0/1/2}^{\mathrm{p/a}}$ are defined as in \citet{Rampp2002} and are taken into account up to 2nd (1st) order for $e^\pm$-annihilation (bremsstrahlung). They are computed using hydro variables at the old time step. In order to circumvent expensive matrix inversions, we also treat the neutrino moments in Eqs.~\eqref{eq:srcpair} explicitly in time for densities below $10^{13}\,$g\,cm$^{-3}$. Above this density, we employ the simplified pair-process treatment (that is used for ``pp'' models at all densities; cf. Sec.~\ref{sec:neutr-inter-invest}), i.e. we ignore pair processes for electron-type neutrinos and assume that annihilation partners for $\nu_x$ neutrinos are in isotropic LTE. This allows to write the pair-process rates formally equivalent to those of emission/absorption processes, cf. Eqs.~\eqref{eq:srcemab}, and to treat the rates implicitly in the neutrino moments without needing to perform matrix inversions. The simplification is justified by the circumstances that at high densities, $\rho>10^{13}\,$g\,cm$^{-3}$, charge-current reactions strongly dominate pair-processes for electron-type neutrinos, and $\nu_x$ neutrinos should be sufficiently close to isotropic LTE.

We have verified that the quenching of neutrino-electron scattering (as described above) as well as the simplification of pair processes at high densities has no significant impact on the results discussed in this paper.

\end{document}